\documentclass[aip,jcp,amssymb,twocolumn]{revtex4}
\pdfoutput=1
\usepackage{amsfonts,amsmath,amsbsy,color,enumitem}
\usepackage{hyperref}
\usepackage[capitalise]{cleveref}
\usepackage[]{graphics,graphicx}
\usepackage[hang,nooneline]{subfigure}
\newcommand{\sr}[1] {_\textrm{#1}}
\newcommand{\order}[1]{{{\cal O}(#1)}}
\newcommand{\D}[1] {D_{\bf #1}}
\newcommand{\ah}[2]{\hat{a}_{#1}^{#2}}
\newcommand{\ahb}[1]{\hat{a}_{\bf #1}}
\newcommand{\braket}[3] {{\langle #1 | #2 | #3 \rangle}}
\newcommand{\brket}[2] {{\langle #1 | #2 \rangle}}
\newcommand{\bket}[1] {{\langle #1 \rangle}}
\newcommand{\ket}[1] {{| #1 \rangle}}
\newcommand{\tb}[1] {t_{\bf #1}}
\newcommand{\figref}[1] {Fig. \ref{#1}}
\newcommand{\ttt}[1] {$\times 10^{#1}$}
\def\sgn{\textrm{sgn}}
\usepackage[version=3]{mhchem}
\crefname{secinapp}{appendix}{appendices}
\Crefname{secinapp}{Appendix}{Appendices}

\begin{document}
\title{Developments in Stochastic Coupled Cluster Theory: The initiator approximation and application to the Uniform Electron Gas}
\author{James~S.~Spencer}
\affiliation{Department of Physics, Imperial College London, Exhibition Road, London, SW7 2AZ, United Kingdom}
\affiliation{Department of Materials, Imperial College London, Exhibition Road, London, SW7 2AZ, United Kingdom}
\author{Alex~J.~W.~Thom}
\email{ajwt3@cam.ac.uk}
\affiliation{University Chemical Laboratory, Lensfield Road, Cambridge CB2 1EW, United Kingdom}
\affiliation{Department of Chemistry, Imperial College London, Exhibition Road, London, SW7 2AZ, United Kingdom}
\date{\today}
\begin{abstract}
We describe further details of the Stochastic Coupled Cluster method and a diagnostic of such calculations, the shoulder height, akin to the plateau found in Full Configuration Interaction Quantum Monte Carlo. We describe an initiator modification to Stochastic Coupled Cluster Theory and show that initiator calculations can be extrapolated to the unbiased limit.  We apply this method to the 3D 14-electron uniform electron gas and present complete basis set limit values of the CCSD and previously unattainable CCSDT correlation energies for up to $r_s=2$, showing a requirement to include triple excitations to accurately calculate energies at high densities.
\end{abstract}
\maketitle
\section{Introduction}
In a recent letter\cite{Thom_10PRL}, we described a novel way of formulating Coupled Cluster theory stochastically.
The previous exposition was limited to calculations on a single processor, but was shown capable, within arbitrarily small error bars, of reproducing exact Coupled Cluster results for a range of molecular systems while requiring lesser computational resources than the full exact calculations.

Coupled Cluster theory, though born in the field of nuclear quantum mechanics, has become astoundingly successful in the field of quantum chemistry.
After the pioneering work of \v{C}\'{\i}\v{z}ek and Paldus \cite{Cizek_66JCP,CizekPaldus_71IJQC} brought it to the quantum chemists' attention, and its use has grown such as to be the `gold-standard' of quantum chemistry\cite{BartlettMusial_07RMP}.  
Its success stems, in essence, from the ability of CCSD(T) (Coupled Cluster Singles and Doubles with perturbative Triples)\cite{RaghavachariHeadGordon_89JCP} to reproduce energetics and properties of a wide range of molecules to `chemical accuracy', i.e. within 1 kJ mol$^{-1}$, and relative ease of formulation, allowing it to be implemented in a range of widely used computational chemistry software.
It has not, however, become ubiquitous within electronic structure theory owing to a high scaling ($\order{N^7}$ for CCSD(T)\footnote{More formally $\order{N^3M^4}$ where $M$ is the number of basis functions in the system}) with number of electrons $N$.
Numerous local approximations\cite{LiakosNeese_15JCTC,SubotnikHeadGordon_06JCP} have reduced this scaling, even to the level of linear with system size\cite{Schutz_02PCCP}, though these have yet to become widespread, and often the onset of the reduced scaling is only in the regime of large systems.
For greater accuracy one can turn to higher levels of truncation of the theory, with full triples, CCSDT, or full triples and quadruples, CCSDTQ, having scalings $\order{N^8}$ and $\order{N^{10}}$ respectively.
This higher accuracy is at the cost of their being vastly more complicated to implement, and their scaling makes them of little current use beyond benchmarking the energetics of very small molecules.  
It is in going beyond this regime that we believe the novel stochastic Coupled Cluster theory to be of great use owing to both its simplicity of implementation and parallelizability.

Within the paradigm provided by stochastic methods, there has been a great resurgence in their applications to quantum chemical methods.
The highly acclaimed work of Booth, Alavi and co-workers\cite{BoothAlavi_09JCP,BoothAlavi_10JCP} in formulating Full Configuration Interaction Quantum Monte Carlo (FCIQMC) as a stochastic implementation of FCI was the impetus for genesis of the stochastic formulation of Coupled Cluster theory,
and the common elements of the methods allow algorithmic developments to be easily transferred between them.  In particular, we shall focus on the `initiator approximation' of Cleland \textit{et al.}\cite{ClelandAlavi_10JCP,BoothAlavi_11JCP,ClelandAlavi_12JCTC} has allowed vastly larger systems to be studied than conventional diagonalization and both work on new properties (explicit correlation\cite{BoothTew_12JCP}, density matrices\cite{OveryAlavi_14JCP}, forces\cite{ThomasBooth_15JCP}, and excited states\cite{BluntBooth_15PRL,BluntAlavi_15JCP}) and new systems (from the Uniform Electron Gas\cite{ShepherdAlavi_12PRB035111,ShepherdAlavi_12PRB081103} and the Hubbard model\cite{ShepherdSpencer_14PRB}, to real crystalline solids\cite{GrueneisKresse_11JCTC,BoothAlavi_13N}).

In this paper, we elucidate more algorithmic details of the CCMC method, as well the inclusion of the `initiator approximation' based upon an analogous development within FCIQMC\cite{ClelandAlavi_10JCP}.
These advances allows the application of high levels of Coupled Cluster theory to problems much larger than previously possible with full Coupled Cluster theory.
In \cref{sec:CCMC} we describe the basic algorithm in detail and provide examples of the Monte Carlo steps involved and denote this algorithm Coupled Cluster Monte Carlo (CCMC).  The population dynamics of CCMC are explored in \cref{sec:population_dynamics}, followed by more details about the critical plateau height in \cref{sec:plateau} in which we propose ``shoulder plots'' as a measure of the critical difficulty of a system, and show how this varies in some molecular systems.   \cref{sec:initiator} introduces the initiator approximation for CCMC and \cref{sec:extrapolation} details how the systematic error inherent to the initiator approximation can be extrapolated to zero.  The behaviour of the CCMC method is explored in these sections using the neon atom and nitrogen molecule, going to six-fold excitations (i.e. CCSDTQ56).  The use of quantum chemistry methods in extended systems has been the subject of much recent interest\cite{ManbyGillan_06PCCP,MarsmanKresse_09JCP,Grueneis_15JCP,DelBenVandeVondele_15JCP,Grueneis_15PRL} and as a further demonstration of the method, we calculate the CCSD and CCSDT energies of the uniform electron gas in \cref{sec:results}.  We find a substantial impact from triple excitations at realistic densities.  Finally we offer some concluding remarks and future perspective in \cref{sec:conclusion}.

\section{Coupled Cluster Monte Carlo}
\label{sec:CCMC}
Key to the success of Monte Carlo techniques is the ability to reduce computational effort by converting sums and integrals over extremely large spaces into a series of discrete samples which approximate the full calculation to arbitrary accuracy with increasing numbers of samples.  
In this section we cast the Coupled Cluster equations in such a form as can be sampled with Monte Carlo techniques, and show how by parameterization as discrete objects in excitation space the Coupled Cluster equations may be easily approximated and solved.

The space in which we shall represent single reference Coupled Cluster theory is that of {\em excitors} which act with respect to a {\em reference} Slater determinant.
Given an orthonormal set of $2M$ spin-orbitals, we partition them into a set of $N$ which are occupied in the reference determinant, denoted ${\phi_i, \phi_j, \dots \phi_n}$, and $2M-N$ unoccupied, or virtual orbitals, denoted ${\phi_a, \phi_b, \dots \phi_f}$.
The complete space of $N$-electron Slater determinants in this basis has size $2M \choose N$ and will be denoted by $\D{m}$ where $\bf m$ is an $N$-vector listing the orbitals occupied in a given determinant. 
Given the occupied/virtual partitioning, we may also represent all possible determinants with respect the the reference (which we will denote $D_0$) by listing the occupied orbitals removed and virtual orbitals added to the determinant, such that $D_{ij}^{ab}$ represents the determinant where $i$ and $j$ in the reference have been replaced by $a$ and $b$.
Further we shall denote by $\ah{ij}{ab}$ the {\em excitation operator} or {\em excitor} which performs this process, $D_{ij}^{ab}=\ah{ij}{ab}D_0$.
The excitors behave such that it is not possible to excite from or to an orbital multiple times, e.g. $\ah{i}{a}\ah{i}{b}=0$ and $\ah{i}{a}\ah{j}{a}=0$.
By applying each of the $2M \choose N$ possible excitors to the reference determinant, the whole of determinant space may be generated, and there is a one-to-one correspondence between excitors and determinants;  we may therefore denote the excitors by the determinant they would create, $\D{i}=\ahb{i}D_0$.

These excitors can be used to parameterize all possible $N$-particle wavefunctions in this basis in a number of ways.  
A simple example would be $\Psi\sr{FCI}=\sum_{\bf i} c_{\bf i} \ahb{i} D_0$, which would correspond a Full Configuration interaction type wavefunction where each determinant has its own coefficient in the expansion.  
Coupled Cluster theory uses instead an exponential {\em Ansatz} for the wavefunction, $\Psi\sr{CC}=e^{\hat{T}}D_0$, where we define $\hat{T}=\sum_{\bf i} \tb{i} \ahb{i}$.
This seemingly complicated parameterization is used owing to its desirable property of remaining size-consistent even if the sum of excitors is restricted to a limited level of excitation.
In essence this is due to the fact that despite a truncation, the exponential ensures that the wavefunction can contain contributions from determinants at all excitation levels.  

To determine the parameters $\{\tb{i}\}$, the projected Schr\"odinger equation is solved; i.e. for all $\bf m$, we solve
\begin{equation}
\braket{\D{m}}{\hat{H}-E}{\Psi\sr{CC}}=0.\label{eq-CC}
\end{equation}
  In general this may be expressed in an iterative form, beginning with a guess for all $\{\tb{i}\}$ and $E$, and iterating until convergence.
The complexity in this arises from the expansion of the exponential
\begin{widetext}
\begin{equation}
e^{\hat T}D_0=\left[1+\sum_{\bf i} \tb{i} \ahb{i}+\frac{1}{2}\sum_{\bf ij} \tb{i} \tb{j} \ahb{i}\ahb{j}+\frac{1}{3!}\sum_{\bf ijk}\tb{i}\tb{j}\tb{k} \ahb{i}\ahb{j}\ahb{k}+\dots\right]D_0.\label{eq-expT}
\end{equation} 
\end{widetext}

Instead of explicitly rearranging the equations \eqref{eq-CC} (or the equivalent more convenient equations, $\braket{\D{m}}{e^{-\hat{T}}(\hat{H}-E)}{\Psi\sr{CC}}=0$) in an iterative form, which is the means by which many conventional implementations work), we note that solutions to the Coupled Cluster equations must also satisfy 
\begin{equation}
\braket{\D{m}}{1-\delta\tau(\hat{H}-E)}{\Psi\sr{CC}}=\brket{\D{m}}{\Psi\sr{CC}},\label{eq-projCC}
\end{equation}

 where $\delta\tau$ is some small positive number\footnote{We have adopted $\delta\tau$ rather than $\tau$ as in \cite{Thom_10PRL} so as to be more consistent with the notation of Alavi {\em et al.}}, giving a form reminiscent of projector methods. 
This is now almost in the form of an iterative method: 
\begin{equation}
\brket{\D{m}}{\Psi\sr{CC}}-\delta\tau\braket{\D{m}}{\hat{H}-E}{\Psi\sr{CC}}=\brket{\D{m}}{\Psi\sr{CC}},\label{eq-projCC2}
\end{equation}
except that the right hand side has components from many different $\tb{i}$ amplitudes, so it is not clear how to perform the iteration.
Noting that the projection of $\Psi\sr{CC}$ on to a single determinant always contains the amplitude of the excitor for that determinant plus terms involving multiple amplitudes, 
\begin{equation}
\brket{\D{m}}{e^{\hat{T}}D_0}=\tb{m} + \order{\hat{T}^2},
\end{equation}
we may cancel out the higher order terms identically on the left and right hand sides of \eqref{eq-projCC2}, leaving
\begin{equation}
\tb{m}-\delta\tau\braket{\D{m}}{\hat{H}-E}{\Psi\sr{CC}}=\tb{m}.
\end{equation}
We may write this as an iteration from time $\tau$ to $\tau+\delta\tau$,
\begin{equation}
\tb{m}(\tau)-\delta\tau\braket{\D{m}}{\hat{H}-E}{\Psi\sr{CC}(\tau)}=\tb{m}(\tau+\delta\tau).\label{eq-CCiter}
\end{equation}
The solutions to the Coupled Cluster equations will also be solutions of this iteration procedure\footnote{The converse, that solutions to the iteration will also be Coupled Cluster solutions, is not necessarily true, though we have not found any situations where this is the case.  Indeed there is no guarantee that this iteration procedure will converge, though with a sensible choice of $\delta\tau$ it appears to be convergent for primarily single-reference systems}, though the evaluation of the second term in \eqref{eq-CCiter} is not trivial.
Indeed, it turns out to be easier to sample this term.  
The most straightforward way of sampling these equations involves storing all $\tb{i}$ amplitudes as real numbers, and performing two stochastic processes. 

The first is sampling $\Psi\sr{CC}$, by selecting randomly from all possible clusters in \eqref{eq-expT}, e.g. 
$\tb{i} \tb{j} \ahb{i}\ahb{j}D_0$.
Once selected the cluster is collapsed to form determinant, $\tb{i} \tb{j} \ahb{i}\ahb{j}D_0=\tb{i}\tb{j}\ahb{n}D_0=\tb{i}\tb{j}\D{n}$.
This process may involve some sign changes or even result in zero (if $\bf i$ and $\bf j$ excite from the same occupied or to the same virtual orbitals).

The second stochastic process is the sampling the action of the Hamiltonian.
For each $\D{n}$ generated from sampling $\Psi\sr{CC}$, we could enumerate all possible $\D{m}$, evaluating $\braket{\D{m}}{\hat{H}-E}{\D{n}}$ and updating $\tb{m}$ accordingly.
As the Hamiltonian only connects up to single and double excitations from a determinant, we may instead sample this process by randomly picking $\D{m}$ as a single or double excitation of $\D{n}$, and updating $\tb{m}$ appropriately.  
While we have found such an algorithm which stores all $\tb{n}$ to well reproduce small Coupled Cluster calculations, it is not efficient, as many clusters selected have extremely small amplitudes, requiring large numbers of samples.

A more efficient route is to {\em discretize} the amplitudes as has been done in the FCIQMC method\cite{BoothAlavi_09JCP}.
In analogy to Anderson\cite{Anderson_76JCP}, we shall represent amplitudes of excitors by sets of {\em excitor particles} or {\em excips}\footnote{Note that our previous publication used the word excitor for both the operator and the particle.}.
Each excip represents a unit weight and is given a positive or negative sign, and located at an excitor, $\ahb{n}$. As the simulation proceeds, excips are created or destroyed at excitors according to some simple rules,
and the mean (averaged over a number of iterations) signed number of excips at a given excitor will be taken to represent its (unnormalized) amplitude.

To sample the action of \eqref{eq-CCiter}, we make very similar steps to the algorithm using real numbers just described.
Firstly a cluster out of $\Psi\sr{CC}$ is selected randomly, and collapsed into a determinant $\D{n}$:  since $\hat{T}$ is represented by populations of excips at different excitors, a cluster is formed by randomly selecting a number of excips (which could be zero).
Clusters in $\Psi\sr{CC}$ formed from excitors with no excips at them would have no amplitude, and so may be safely ignored.
The cluster is collapsed by taking the product of the excitors at which each of the chosen excips is located to give determinant $\D{n}$.
The amplitude of the cluster is given by the products of populations of excips at each of the excitors in the cluster.
  As the spawning probability must be proportional to this, we have chosen to select each cluster with a probability proportional to its amplitude, and make one spawning attempt per cluster.
Then, much as in FCIQMC, we follow the following steps:
\begin{enumerate}
\item {\em Spawning}.  From each $\D{n}$ generated we pick a random connected single or double excitation, $\D{m}$, and create a new excip there with probability proportional to $|\delta\tau H_{\bf mn}|$ and appropriate sign.
\item {\em Birth/Death}. From each $\D{n}$ generated, with probability proportional to $|\delta\tau(H_{\bf nn}-E)|$ we create an opposite signed excip at $\ahb{n}$.
Here $E$ has yet to be defined, but may be taken as an arbitrary constant.
Its later role will be to control population growth.
  Because $\D{n}$ may have been formed from a product of excitors, it is probable that there are no excips at the excitor $\ahb{n}$, and so, in contrast to FCIQMC,  we must create an oppositely signed excip, rather than simply killing an existing one.
If the cluster consisted of a single excip, the opposite signed excip will cancel this out.
\item {\em Annihilation}.  Finally, after these two processes have been done for as many clusters as required to be sampled, the new list of excips is sorted and any opposite signed pairs of excips on the same excitor removed.
\end{enumerate}
In deterministic Coupled Cluster theory, $E$ is a parameter which iteratively converges to the Coupled Cluster energy. Here we shall replace it with a parameter $S$ which will shift the diagonal elements of the Hamiltonian and may be updated periodically after a number of cycles to allow the total population of excips to be controlled\cite{BoothAlavi_09JCP,UmrigarRunge_93JCP}.

\subsection{Normalization}
The careful reader may have noticed that above formalism introduces an imbalance in the description of the reference and the excitor space.
Indeed, the prescription of intermediate normalization, where the overlap of the reference with the Coupled Cluster wavefunction is fixed at unity, makes it rather difficult to satisfy \eqref{eq-projCC2} for $\D{m}=D_0$ in terms of an iteration.
Instead, it is convenient to introduce a further variable, $N_0$, to act as the normalization constant,
\begin{equation}
\ket{\Psi\sr{CC}}=N_0 e^{\frac{\hat{T}}{N_0}}\ket{D_0}.
\end{equation}
The inclusion of $N_0$ within the exponential allows the discretization of $\hat{T}$ to produce effectively fractional populations on excitors which is essential for the exponential to converge.
With this normalization, the number of excips at the reference is given by $N_0$, and can thus be varied in the iteration satisfying \eqref{eq-projCC2}. 
In this manuscript we will refer to these particles on the reference as excips, though they are not strictly the same, and cannot be used as a constituent excip of a cluster; as such they are (all) always in a cluster as they normalize it multiplicatively.

\subsection{Energy estimators}
There are two independent estimators of the Coupled Cluster energy available in this formalism.
Firstly, the time-averaged value of the shift is, at convergence, an estimator for the energy.
Secondly the energy may be estimated by projection as
\begin{equation}
E\sr{proj}=\frac{\braket{D_0}{\hat{H}}{\Psi\sr{CC}}}{\brket{D_0}{\Psi\sr{CC}}}\label{eq-projE}
\end{equation}
This expression could be computed exactly, as the contributions from only singly and doubly excited determinants are relevant, but would require an iteration through all possible pairs of single excitors to evaluate their contribution to the doubly excited determinants.
Instead, we sample both of these contributions while sampling the cluster wavefunction.

\subsection{Sampling}
Sampling forms the core of this method, and the implementational details of this are crucial to an efficient calculation.
Here we describe the different sampling schemes used in our implementation.
Though we have not performed exhaustive tests on different sampling methods, these have appeared the most effective in our calculations.

\begin{enumerate}
\item {\em Sampling the wavefunction}.  At each timestep we must sample the exponential expansion of the wavefunction, and we do this by randomly selecting a cluster --- a single term consisting of a product of specific amplitudes and excitors --- from the expansion of the exponential.  
We have found that setting the number of samples taken to be the same as the number of excips leads to stable simulations.
If the number of samples does not scale at least linearly with the number of excips, the selection probabilities of the clusters become increasingly small and lead to population blooms which destabilize the simulation.

We select the size of cluster, $s$, to be generated so that the probability of selecting a cluster with $s$ excitors decays exponentially with $s$.
Each cluster is generated by selecting $s$ excips randomly from the list of excips, and we have found it best to bias the selection so that the probability of selecting an excip is proportional to the total number of excips at that excitor.

\item {\em Sampling the spawning}.  In principle the spawning step samples the action of the Hamiltonian upon the collapsed cluster.
A full sampling would enumerate all connected determinants and attempt to spawn on each one-by-one.
We have followed the route of Booth {\em et al.}\cite{BoothAlavi_09JCP} in choosing a minimalist sampling of a single attempted spawn from each cluster.
This comes with the drawback that as the number of connected determinants increases (say with system size or basis), the timestep must be correspondingly reduced to keep this step stable.
Some brief tests of increasing the number of spawnings from each cluster indicate this has the same effect as dividing the timestep by this number (reducing the effective timestep).

\item {\em Sampling the projected energy}.  The explicit evaluation of the numerator of \eqref{eq-projE} is complicated by pairs of single-excitations.
Instead, each time a cluster is generated, its contribution to the numerator is accumulated (appropriately unbiased by its generation probability).
\end{enumerate}

\subsection{A Representative Example}
Let us consider a spin-free system with occupied orbitals $i,j,k$ and virtual orbitals $a,b,c$.
In addition to the reference there are 19 possible excitors, which we shall denote with subscripts and superscripts.  For example, $_{ik}^{ab}$, denotes the excitor which excites electrons from $i$ and $k$ to orbitals $a$ and $b$.

\begin{figure*}
\subfigure[A size-0 cluster spawns.]{\includegraphics[scale=0.8]{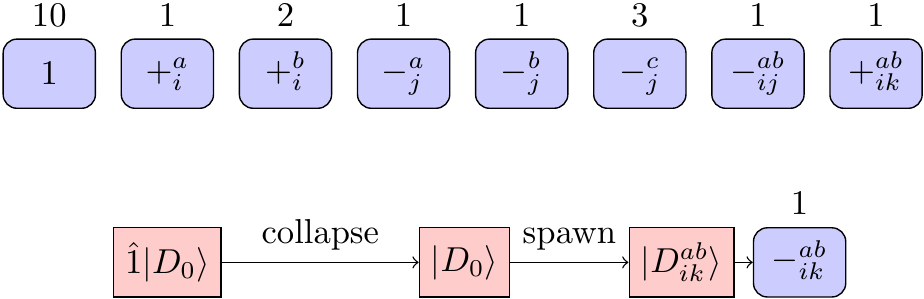}\label{fig-ex-1}}\\

\subfigure[A size-2 cluster spawns.]{\includegraphics[scale=0.8]{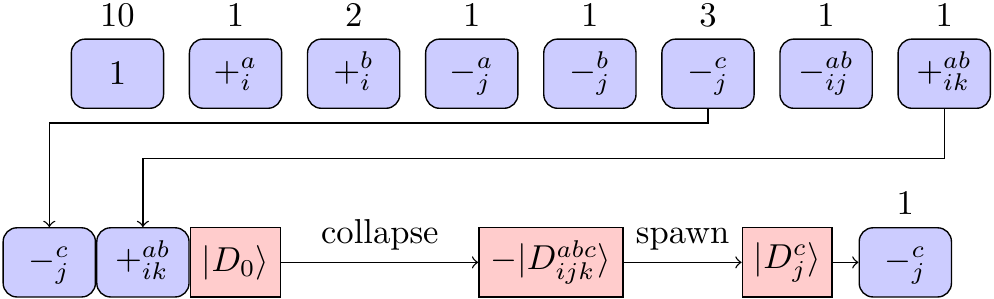}\label{fig-ex-2}}\\

\subfigure[A size-2 cluster spawns and dies.]{\includegraphics[scale=0.8]{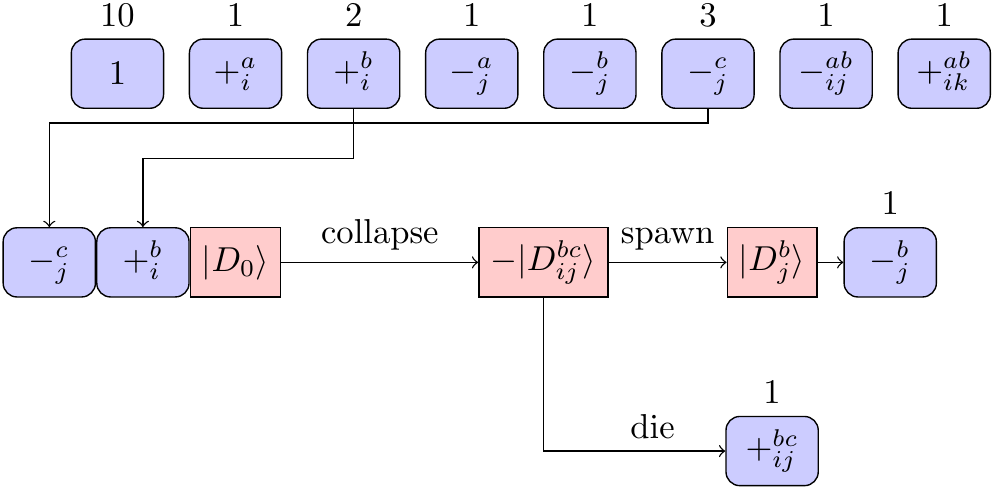}\label{fig-ex-3}}\\

\subfigure[ A size-1 cluster aborts a spawn outside the truncation level and dies.]{\includegraphics[scale=0.8]{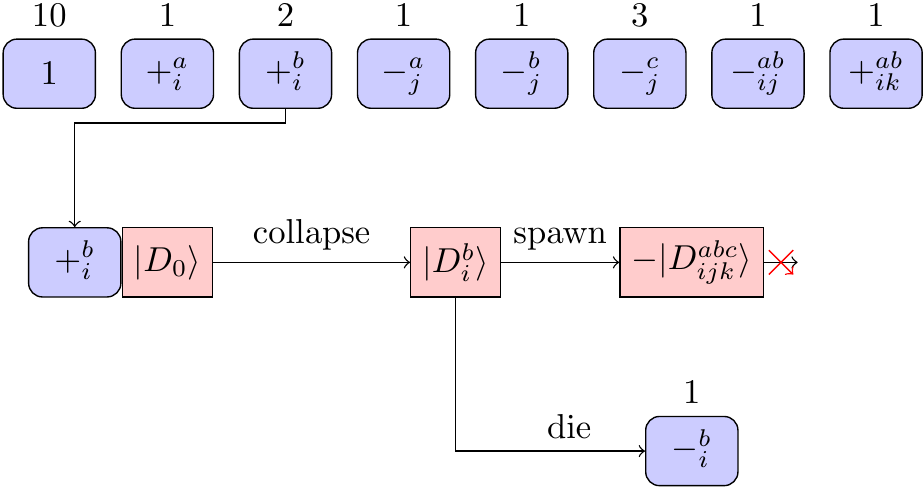}\label{fig-ex-4}}\\

\subfigure[The post-annihilation populations.]{\includegraphics[scale=0.8]{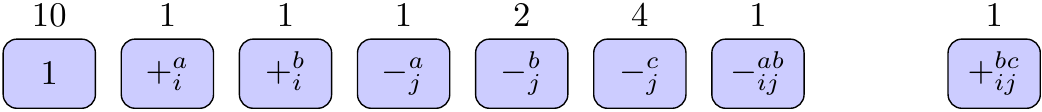}\label{fig-ex-5}}

\caption{A representative example of the cluster selection, spawning and death steps of the CCMC algorithm as explained in the main text.  Rounded blue boxes represent excitors, and rectangular blue boxes represent determinants. }
\label{fig-ex}
\end{figure*}

Let us suppose we are performing Coupled Cluster theory truncated at singles and doubles (a rather futile truncation since there is only one triple excitation), i.e. CCSD.
At time $\tau$ in the simulation, we have reached the situation at the top of \figref{fig-ex-1}, with 20 excips, 10 being at the reference.
We shall accumulate various generation probabilities, $p\sr{X}$, through this example which will be used to unbias our sampling.
We sample with the same number of samples as excips (hence $p\sr{sel}=20$ for this timestep).
We shall also adopt a scheme where the likelihood of forming a cluster is exponentially decaying with probability $\frac{1}{2}$.
Since the largest cluster with any effect is of size three, we stop at that point and add in the remaining probabilities of selection for larger clusters to that level.
Let us also suppose that we have not yet modified the shift from its default $S=E\sr{HF}$.
The following four examples of samples and their evolution are illustrated in \figref{fig-ex-1}--\ref{fig-ex-4}.
\begin{enumerate}[label=(\alph*)]
\item We decide with probability $p\sr{size}=\frac{1}{2}$ to select a cluster with zero excips.
There is only one choice of zero-cluster, so we set $p\sr{clust}=1$.
The action of this on the reference is merely $\hat{1}\ket{D_0}=\ket{D_0}$, and it has total amplitude $A=N_0$. There are eighteen possible single and double excitations available, and we pick $D_{ik}^{ab}$ with $p\sr{excit}=\frac{1}{18}$. $\ah{ik}{ab}\ket{D_0}=\ket{D_{ik}^{ab}}$, so no sign change will be needed.
We create a new excip at $_{ik}^{ab}$ with probability $\delta\tau |\braket{D_{ik}^{ab}}{\hat{H}}{D_0} A|/ (p\sr{sel}p\sr{size}p\sr{clust}p\sr{excit})$, and sign $-\sgn(\braket{D_i^b}{\hat{H}}{D_0})$.
Let's say we create a negative excip at $_{ik}^{ab}$.
 Death occurs with probability $|(\braket{D_0}{\hat{H}}{D_0}-S) \delta\tau A|/ (p\sr{sel}p\sr{size}p\sr{clust})$. Here as $E\sr{HF}=S$, there is no death.
\item We decide with probability $p\sr{size}=\frac{1}{8}$ to pick a cluster of size 2.
$-_j^c$ is selected with probability $p\sr{1}=\frac{3}{10}$ and $+_{ik}^{ab}$ is selected with probability $p\sr{2}=\frac{1}{10}$, so $p\sr{clust}=(2!)p\sr{1}p\sr{2}$, the $2!$ accounting for the number of ways in which the cluster could have been chosen. The amplitude of the cluster is given by $A=N_0\frac{-3}{N_0}\frac{1}{N_0}$.
The cluster $\ah{j}{c}\ah{ik}{ab}$ collapses to $-\ket{D_{ijk}^{abc}}$ when applied to the reference. We first spawn, picking $\ket{D_j^c}$ with probability $p\sr{excit}=\frac{1}{18}$.
$\ah{j}{c}\ket{D_0}=-\ket{D_j^c}$ so we will pick up an extra negative if we spawn there, which we do with probability $\delta\tau |\braket{D_j^c}{\hat{H}}{D_{ijk}^{abc}} A|/(p\sr{size}p\sr{clust}p\sr{excit})$. The signs take a little more care.
Let's say that $\braket{D_j^b}{\hat{H}}{D_{ijk}^{abc}}$ is negative, giving a collective five negatives, resulting at a negative excip at $_j^c$.
Death would occur with probability $|(\braket{D_{ijk}^{abc}}{\hat{H}}{D_{ijk}^{abc}}-S) \delta\tau A|/ (p\sr{sel}p\sr{size}p\sr{clust})$, except that in CCSD, we do not store excips for triples, so it is ignored.
\item We decide with probability $p\sr{size}=\frac{1}{8}$ to pick a cluster of size 2.
$-_j^c$ is selected with probability $p\sr{1}=\frac{3}{10}$ and $+_i^b$ is selected with probability $p\sr{2}=\frac{2}{10}$, giving $p\sr{clust}=(2!)p\sr{1}p\sr{2}$.
The amplitude is $A=N_0\frac{-3}{N_0}\frac{2}{N_0}$. The cluster $\ah{j}{c}\ah{i}{b}$ collapses to $\ket{D_{ij}^{bc}}$ when applied to the reference. We pick $\ket{D_j^b}$ with probability $p\sr{excit}=\frac{1}{18}$.
$\ah{j}{b}\ket{D_0}=-\ket{D_j^b}$ so we will pick up an extra negative for spawning, which has probability $\delta\tau |\braket{D_j^b}{\hat{H}}{D_{ij}^{bc}} A|/(p\sr{size}p\sr{clust}p\sr{excit})$. A positive value of $\braket{D_j^b}{\hat{H}}{D_{ij}^{bc}}$ results in a negative excip being spawned at $_j^b$. Death occurs with probability $|(\braket{D_{ij}^{bc}}{\hat{H}}{D_{ij}^{bc}}-S) \delta\tau A|/ (p\sr{sel}p\sr{size}p\sr{clust})$, creating a positive excip at $_{ij}^{bc}$ because $A$ is negative. 
\item We decide with probability $p\sr{size}=\frac{1}{4}$ to pick a cluster of size 1.
$+_i^b$ is selected with probability $p\sr{1}=\frac{2}{10}$, giving $p\sr{clust}=(1!)p\sr{1}$.
The amplitude is $A=N_0\frac{2}{N_0}$. The cluster $\ah{i}{b}$ collapses to $\ket{D_{i}^{b}}$ when applied to the reference. We pick spawning excitation $\ket{D_{ijk}^{abc}}$ with probability $p\sr{excit}=\frac{1}{18}$.  This is a triple excitation and outside our truncation so it is discarded.
Death occurs with probability $|(\braket{D_{i}^{b}}{\hat{H}}{D_{i}^{b}}-S) \delta\tau A|/ (p\sr{sel}p\sr{size}p\sr{clust})$, creating a negative excip at $_{i}^{b}$. 
\end{enumerate}
Let us assume that all the other samplings result in no death or spawning.
The resultant list of new excips is merged with the main list, annihilation occurs, and a new step begins (\figref{fig-ex-5}).
We note that in this implementation much of the amplitude information cancels between $A$ and $p\sr{clust}$, resulting an a relatively efficient sampling.

\section{Population Dynamics}
\label{sec:population_dynamics}

\begin{figure}
\subfigure[ ]{\includegraphics[scale=0.51]{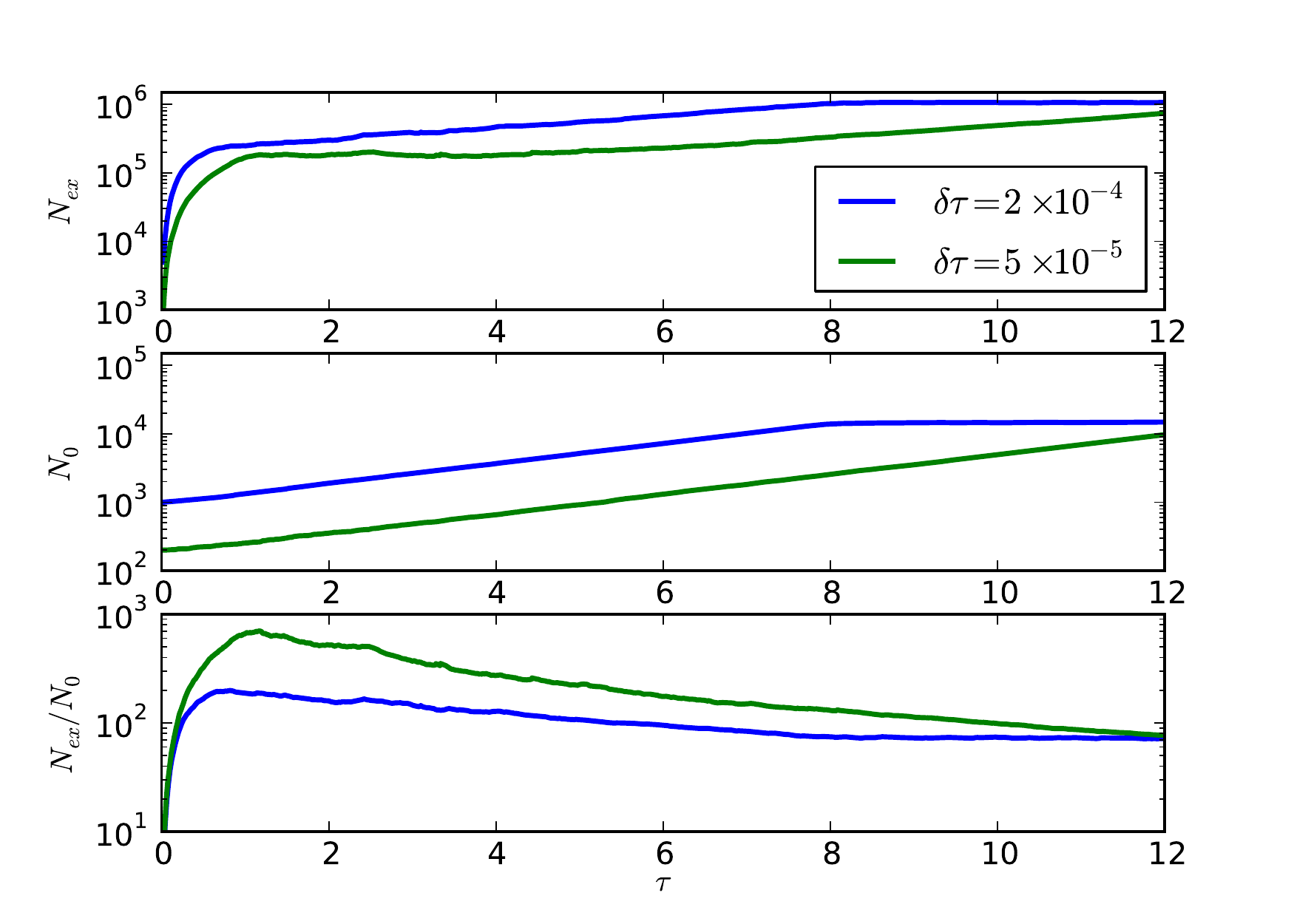}\label{fig-shoulder-1}}
\subfigure[ ]{\includegraphics[scale=0.51]{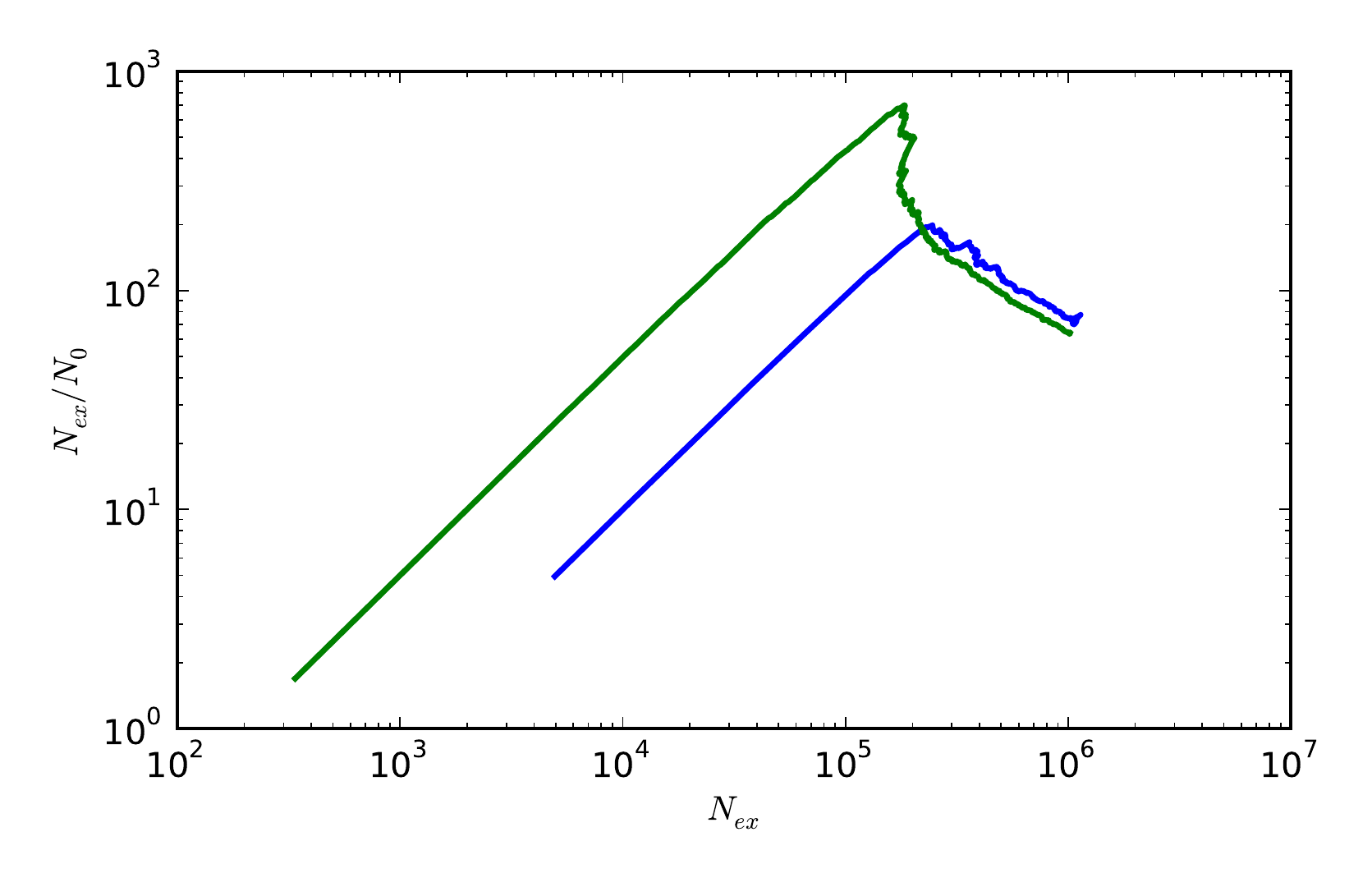}\label{fig-shoulder-2}}
\caption{Ne cc-pVQZ CCSDTQ calculations starting with different initial particle numbers at the reference and different timesteps. (a):  With a carefully chosen low timestep and initial population, a plateau is visible.  An increased timestep and initial population overshoots the plateau but has a shoulder.  The lower panel shows a maximum of the particle ratio at the position of the shoulder and plateau. (b): `Shoulder plots' allow shoulder height to be read off easily and calculations compared.}
\label{fig-shoulder}
\end{figure}

The behaviour of the such algorithms has been previously explored in the context of FCIQMC\cite{SpencerFoulkes_12JCP}, and follows the following phases as shown in Fig \ref{fig-shoulder}:
\begin{enumerate}
\item {\em Initial Growth}. We beginning with a population of excips at the reference. By initially setting $S=E\sr{HF}$ we ensure that the excips at the reference do not die. Gradually, the excitor space is populated with excips spreading outwards from the reference.
As time progresses and the populations grow, annihilation events become more common. 
\item {\em Annihilation Plateau}.  At a system-dependent number of excips, annihilation events become equal to spawning events, and the population ceases to grow, but remains approximately constant.
The relative signs of the populations at excitors where there are many excips changes over this time until it resembles the sign structure appropriate to the ground state wavefunction (as parameterized by excitation amplitudes).
\item {\em Exponential Growth}.  The sign structure of the excips is such that spawning once again dominates over annihilation, and the total number of excips grows exponentially with growth rate proportional to $E-E\sr{CC}$, where $E\sr{CC}$ is the Coupled Cluster energy.
\item {\em Population Control}.  $S$ is now dynamically modified to control the population and limit growth, with more negative values of $S$ reducing population growth.
\end{enumerate}

\begin{figure}
\includegraphics[scale=0.51]{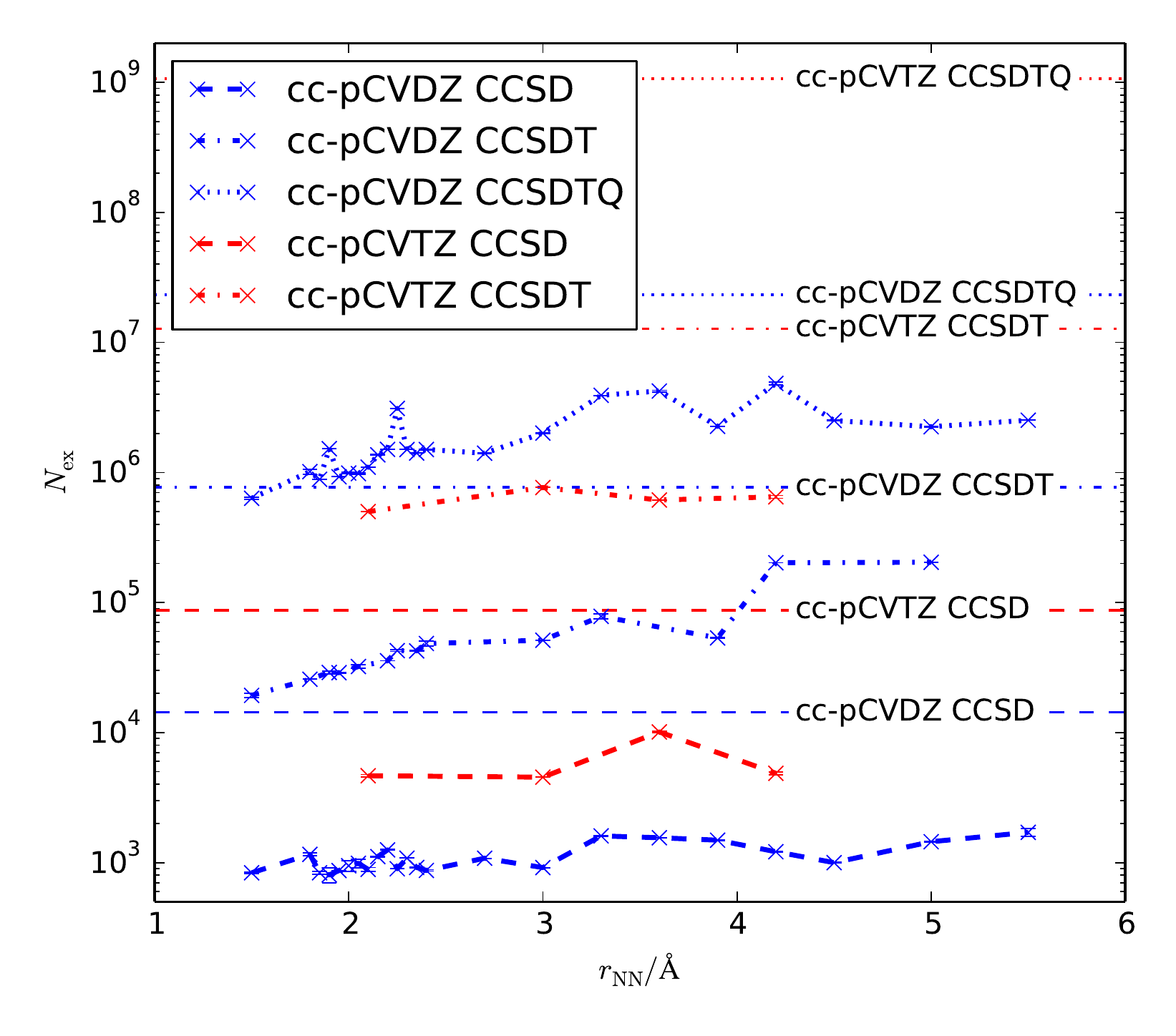}
\caption{Shoulder heights for the \ce{N2} molecule at different bond lengths, basis sets and truncations.  All electrons were correlated, and $L_z$ symmetry enforced.  Hilbert space sizes are shown by the horizontal lines.}
\label{fig-NNplateau}
\end{figure}

\begin{figure}
\includegraphics[scale=0.51]{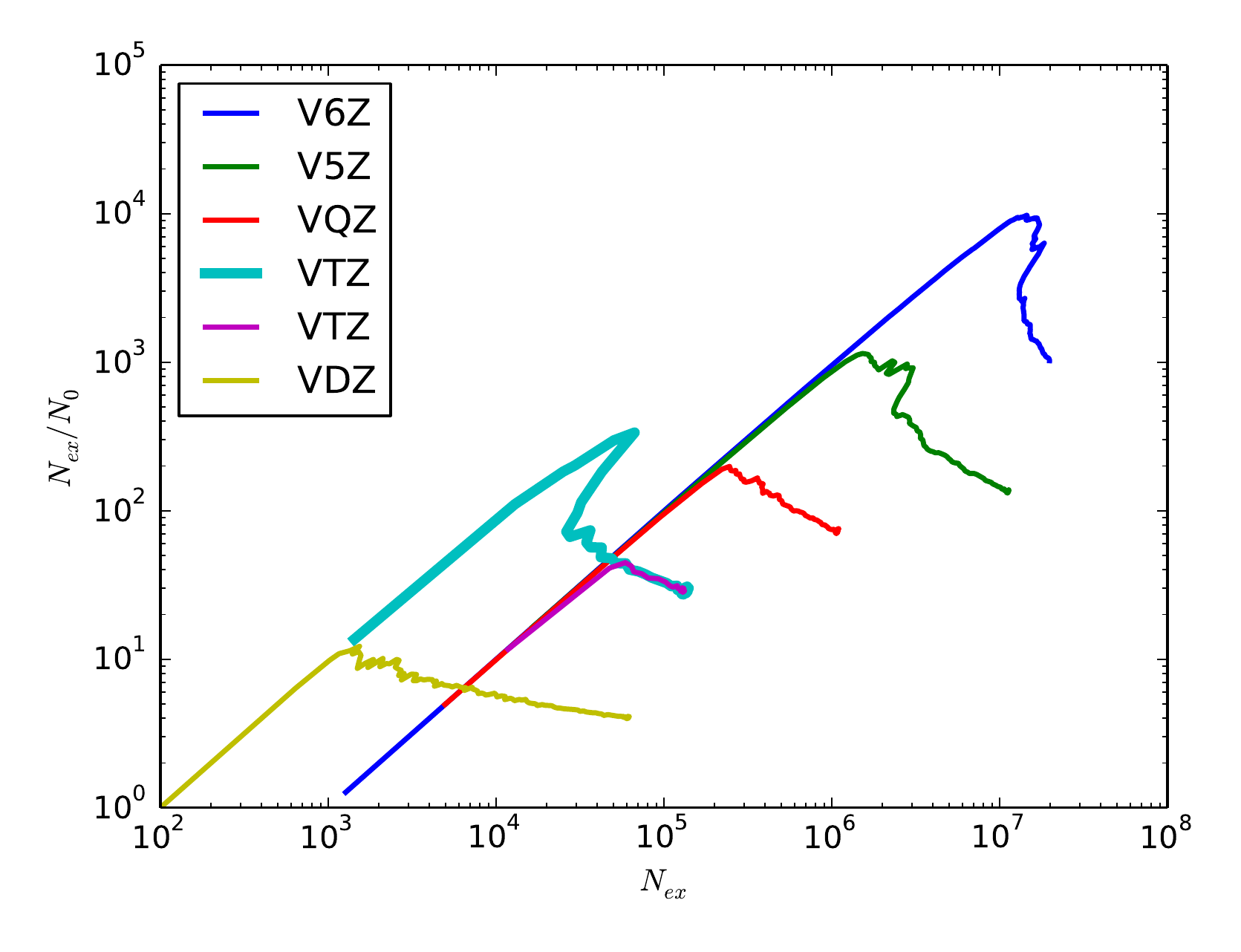}
\caption{All-electron CCSDTQ calculations on the neon atom using Dunning's cc-pVXZ basis sets\cite{Dunning_89JCP,WilsonDunning_97JMS}.  The shoulder is clearly discernible in each line, and its height increases with basis set cardinality.  A vertical descent from the maximum is given by a clearly defined plateau, and a chevron-like peak is characteristic of a shoulder.  As seen for cc-pVTZ, calculations started with more excips lead to a shoulder which can overestimate the position of a plateau.  Calculations were started with $10^2$ or $10^3$ excips at the reference (where each line would cross $N\sr{ex}/N\sr{0}=10^0$).}
\label{fig-shoulderNe}
\end{figure}

Before the Annihilation Plateau, the total rate of growth is faster than that at the reference, and if one attempts to control the population growth with the shift, the population at the reference decreases to zero, causing the simulation to fail.
Below the plateau, in common with FCIQMC, we also expect the sign structure of the solution to be wrong\cite{SpencerFoulkes_12JCP}, and so the plateau height becomes a crucial feature in the usability of Stochastic Coupled Cluster.

We note that, as we begin simulations with a relatively large population of excips at the reference, often the rate of annihilation does not exactly match that of growth and the population growth moves through a shoulder rather than a plateau (see figure \ref{fig-shoulder}).
This is caused by effective overshooting the plateau phase by starting with more excips.
To determine whether the system has reached the region of exponential growth, we note that in this region, the population of at all excitors should grow exponentially.
A suitable alternative is therefore to check whether the population at the reference is growing at the same rate as that of the total population.
One this has been satisfied, the shift, $S$, may be allowed to vary to control the population.

\section{The Plateau}
\label{sec:plateau}
The stability of a CCMC calculation requires reaching a plateau or shoulder in the particle growth.
Discerning such a feature by eye is not an easy task, so we sought a more rigorous definition.
Before the plateau, a shift based on the (faster) total particle growth rate would cause a decrease in the population at the reference, and likely lead the calculation to fail as this population tends to zero.
The post-plateau stability arises from the fact that the growth rate of reference particles is at least as fast as those of the total particles in the system, and so turning on a shift will slow these rates equally and the populations will stabilize.
On this basis, we suggest that the ratio of total particles to reference particles should be a useful measure of these rates of growth, and we shall call this the \textit{particle ratio}. 
In figure \ref{fig-shoulder} we plot a typical CCMC calculation showing a plateau and a second calculation on the same system with different timestep and larger initial population.
The calculation with larger initial population overshoots the plateau, but still reaches a stable growth phase.
The particle ratio shows clearly the location of the plateau by moving from increasing to downward; the overshooting calculation shows the same behaviour corresponding the shoulder in its growth.

From some experience with comparing such plots, it proves convenient to plot the particle ratio not against the progress in imaginary time, but against the total number of particles, both axes being logarithmic.
In such a plot the plateau appears as a sharp downward kink in an otherwise apparently linear progress.
The usefulness of such `shoulder' plots in comparing different calculations is explained in figure \ref{fig-shoulder-2}.
There is still some potential uncertainty as to the definition of the shoulder height, but from experience it has proven to be useful to define this to be at the maximum of the shoulder plot, and this should be taken as an upper limit of the plateau height of the system.
Such a definition is still subject to stochastic noise, and so in this paper we have quoted values which are the mean of the total excip populations with the ten largest particle ratios, and used the standard deviation of these as a measure of uncertainty.
Calculations with total particle numbers before the shoulder are often unstable, and calculations after the shoulder are usually stable, though the stochastic nature of such calculations removes total certainty about this.
A stable population (due to the control from the variable energy shift) appears in the shoulder plot as a tight blob at the end of the line due to (comparatively) small stochastic fluctuations in both the population on the reference and the total population.

A histogram approach has been previously used to automate detection of the plateau in FCIQMC\cite{ShepherdSpencer_14PRB}; the shoulder plot is equally amenable to automation and both approaches give similar values for simulations with clear plateaus.  Shoulder plots are more reliable when the plateau is shorter in duration or tends towards a shoulder in the population growth. 

As shown in the original CCMC paper\cite{Thom_10PRL}, the heights of plateaux vary with both the size of Hilbert space and differing levels of correlation within a molecule; system-specific behaviour is also seen for FCIQMC calculations\cite{BoothAlavi_09JCP,ShepherdSpencer_14PRB}.  Figure \ref{fig-NNplateau} shows the variation of plateau height with basis, excitation and bond length for the \ce{N2} molecule, showing broadly that the fraction of the Hilbert space required is of a similar magnitude for different excitation levels and basis sets, and increases as the \ce{N-N} bond is broken, though staying below 27\%.

To see the effects of different basis sets, the neon atom, with only 10 electrons, is more amenable to study, and Figure \ref{fig-shoulderNe} shows examples of shoulder plots for increasingly large bases.  Plateau heights and correlation energies are given in Table \ref{tab-Neplateaux} and decrease with truncation level, heading to less than 1\% of the complete Hilbert space size for larger bases with quadruples truncation.
Whilst for CCSD, therefore, the saving is relatively small, it is for higher excitation levels that the Monte Carlo sampling considerably reduces the storage required.
\begin{table*}
\begin{tabular}{c|cc|cc|cc|cc|cc}
truncation&\multicolumn{2}{c}{SD} &\multicolumn{2}{c}{SDT}   &\multicolumn{2}{c}{SDTQ}      &\multicolumn{2}{c}{SDTQ5}     &\multicolumn{2}{c}{SDTQ56}    \\
\hline
basis     & H.S. & Shoulder       & H.S.       & Shoulder    & H.S.        & Shoulder       & H.S.        & Shoulder       & H.S.         & Shoulder      \\
cc-pVDZ   & 400  & 33(1)          & 4680       &  194(2)     & 30654       &  594(4)        & 113550      & 1610(20)       & 259460       & 2660(10)      \\
cc-pVTZ   & 1940 & 160(30)        & 60710      &  2900(300)  & 1.032\ttt{6}&  2.27(1)\ttt{4}& 1.021\ttt{7}& 1.90(5)\ttt{4} & 6.143\ttt{7} & 9.6(2)\ttt{5} \\
cc-pVQZ   & 6710 & 290(10)        &4.128\ttt{5}& 7100(200)   & 1.412\ttt{7}&  2.39(6)\ttt{5}& 2.81\ttt{8} & 3.7(1)\ttt{6}  &&\\
cc-pV5Z   & 17100& 2500(400)      &1.815\ttt{6}&7.9(2)\ttt{4}& 1.068\ttt{8}&  1.57(6)\ttt{6}& 3.66\ttt{9} &&&\\
cc-pV6Z   & 82400& 2.2(3)\ttt{4}  &1.636\ttt{7}&2.3(3)\ttt{5}& 1.731\ttt{9}&  1.45(3)\ttt{7}& &&&\\
\end{tabular}
\caption{All-electron Ne Hilbert space (H.S.) sizes and shoulder heights. Hilbert space sizes were determined by Monte Carlo and only significant figures are shown.  $L_z$ symmetry was enforced except in the case of cc-pV6Z. Shoulder error estimates in the last significant digit are shown in parentheses.}
\label{tab-Neplateaux}
\end{table*}

\section{The Initiator Method}
\label{sec:initiator}
\begin{figure}
\includegraphics[scale=0.51]{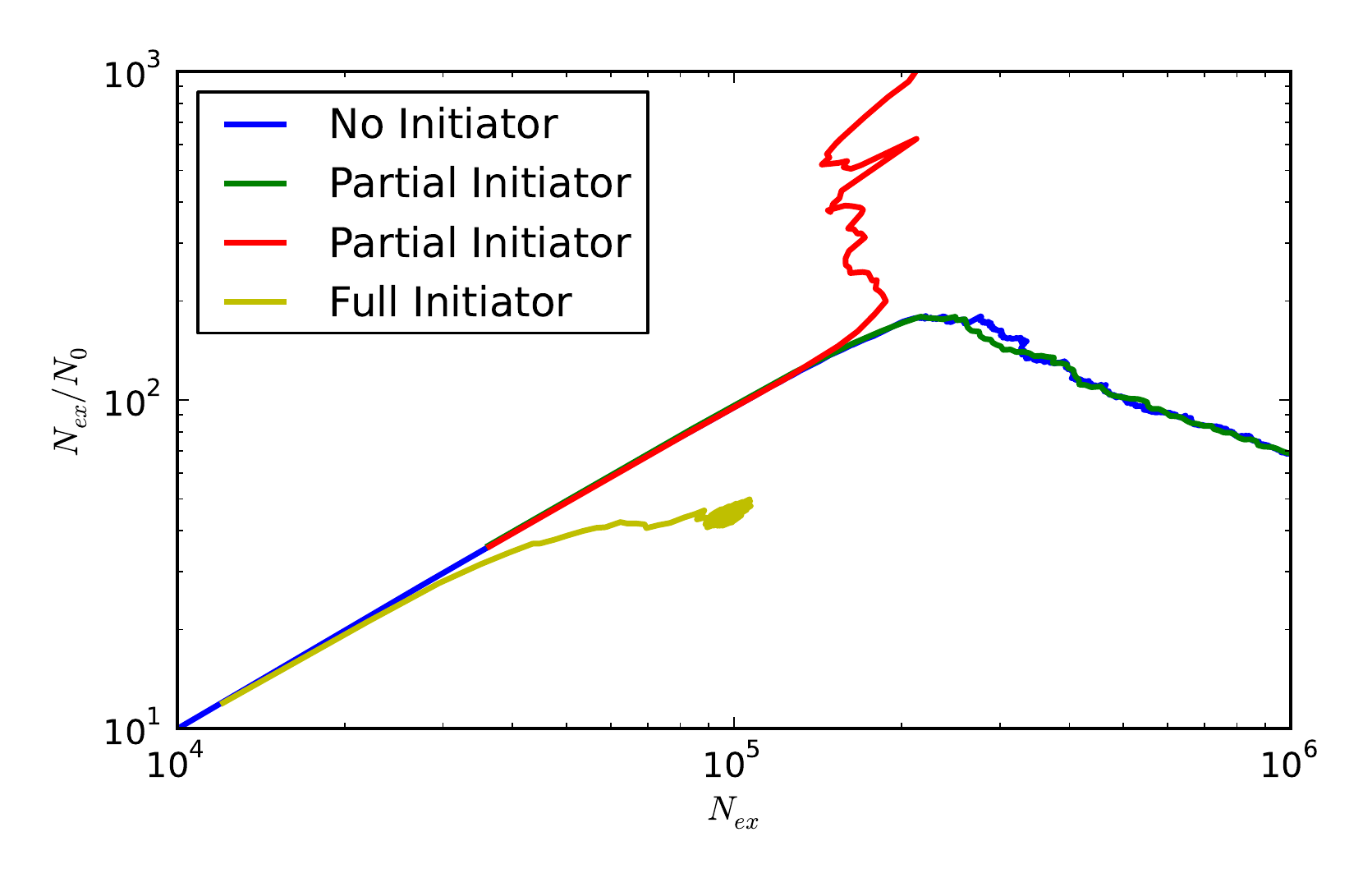}
\caption{Ne cc-pVQZ CCSDTQ calculations without initiator, and with partial and full initiator methods.  The full initiator (yellow) method has a significantly lower shoulder height and is stable at even 100,000 excips whereas the partial (red) is unstable when the shift is engaged at 240,000 excips (before the plateau), and has a similar profile (green) to the non-initiator (blue) method.}
\label{fig-init}
\end{figure}

With the insight that the plateau is exited only when the particles achieve a sign-coherent structure, Cleland et {\em al.} introduced what they call the `initiator method'\cite{ClelandAlavi_10JCP} to the FCIQMC algorithm (denoted i-FCIQMC).
In its mature version it avoids the effects of sign-incoherence by restricting spawning.
If a determinant has no walkers present, then the algorithm only allows a walker to be spawned there if there are more than a critical number, $n\sr{add}$, of walkers present at the determinant from which it was spawned.  
This modification of the algorithm dramatically alters the dynamics, removing the plateau phase and greatly stabilizing the calculations.
This is, however, at the price of introducing a bias into the algorithm, dependent on the number of walkers.  This is manifested as an error in the energy which systematically decreases with the number of walkers (becoming zero in the large walker limit).

Owing to the great similarity between Stochastic Coupled Cluster theory and FCIQMC, it is expected that the initiator method will confer similar benefits in Stochastic Coupled Cluster theory.
As spawning occurs from clusters of excips rather than single walkers, the initiator method requires some modification.
The obvious choices are for a cluster to be allowed to initiate (i.e. spawn onto an unoccupied excitor) if either one or all of its component excips come from an excitor whose population is $>n\sr{add}$.  We denote these the partial and full initiator methods respectively.
A representative sample of the behaviour of each of these schemes is shown in Figure \ref{fig-init}.
From this, and with the ideas of maintaining sign-coherence, we have settled upon requiring all excips within a cluster to come from excitors with populations $>n\sr{add}$ before allowing it to initiate (the full initiator method).

\begin{figure}
\includegraphics[scale=0.51]{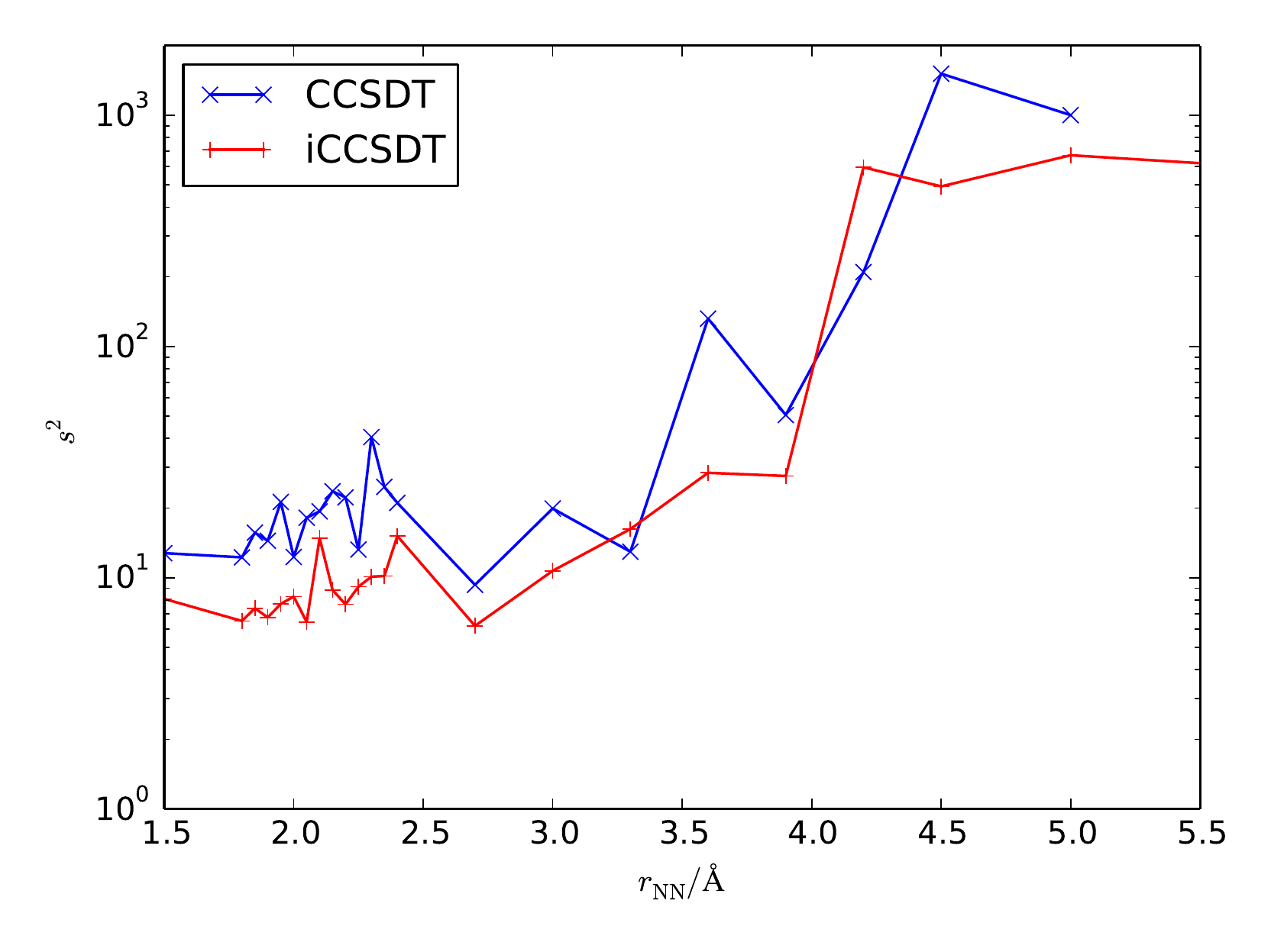}
\caption{The normalized relative variance of the projected energy, $s^2$, (see text for definition) for the \ce{N2} molecule in a cc-pCVDZ basis. Calculations were performed with about 70,000 excips (for $r\sr{NN}\le2.4\,$\AA) and 200,000 otherwise.  The initiator calculations show a generally reduced variance around equilibrium, leading to fewer steps being required to converge to a given error estimate.}
\label{fig-NNerror}
\end{figure}

As the initiator calculation reduces the amount of spawning, it can be thought of as generally reducing the particle noise within a calculation.  To investigate the effect of this, we note that in the regime of large numbers of particles, the variance of the numerator and denominator of the projected energy should decrease with $1/N\sr{ex}$.  More formally, denoting the numerator of the projected energy at a given timestep to be $E\sr{proj,numer}(\tau)$, to enable comparison of different calculations, we may calculate
\begin{equation}
s^2=\frac{\mathrm{Var}[E\sr{proj,numer}(\tau)]_\tau\langle N\sr{ex}(\tau)\rangle_\tau}{\langle E\sr{proj,numer}(\tau)\rangle_\tau^2},
\end{equation}
which we shall call the normalized relative variance of the projected energy.  This is a substantial component of the eventual error estimate for a given calculation (though the actual errors are calculated through a blocking analysis\cite{FlyvbjergPetersen_89JCP}).
A calculation with double the value of $s^2$ will have to be run for approximately twice as long to achieve the same estimate of error, and so it should be as low as possible to reduce the computational resources required in a calculation.
For \ce{N2} in a cc-pCVDZ basis, we show $s^2$ for both standard and initiator calculations in Figure \ref{fig-NNerror}.
Around equilibrium, it can be seen that the variance is significantly lower for initiator calculations, and so using the initiator method should dramatically reduce the number of steps required to achieve a given error bar.

In Figure \ref{fig-Ne-V6Z-x3}, we show data from a typical set of initiator calculations using CCSDT on the neon atom in a cc-pV6Z basis.
As the population increases, the correlation energy found gradually converges to the true correlation energy.
In this case the population required to for the initiator approximation's bias to have decayed is of the order of $5\times10^5$ psips, compared to a shoulder height of $2.3(3)\times10^5$ psips, so providing no benefit in reducing the number of psips needed.
In all initiator CCMC calculations we have tried on systems in this paper (on the neon atom, \ce{N2} molecule, and UEG), this pattern is repeated, and we see that a population of the order of the shoulder height is required to converge the initiator error.
This is in contrast to in FCIQMC where the initiator adaptation substantially helps the sign problem allow much smaller populations to be used, and unfortunately we do not see any similar reductions in CCMC.
The reduction in variance is, however, still beneficial.

\section{Extrapolation}
\label{sec:extrapolation}
\begin{figure}
\includegraphics[scale=0.51]{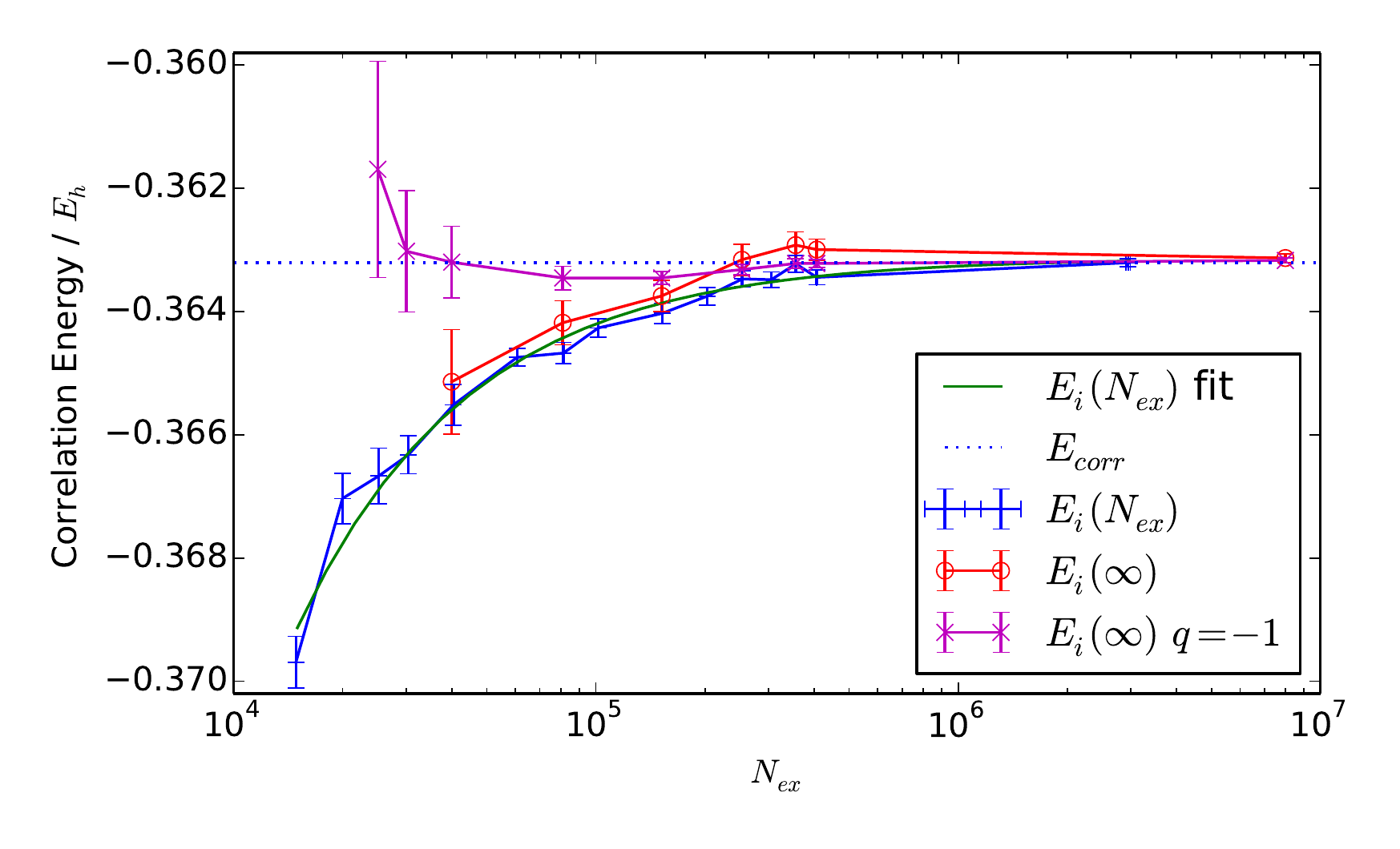}
\caption{CCSDT initiator energies and extrapolations for Ne cc-pV6Z. Plotted are the raw iCCMC energies (blue), $E_i(N_{ex})$, and various fits.  The curve (green) was fit using all data to the form $E_i(N_{ex})=p(N_{ex})^q+E_i(\infty)$, varying $p, q,$ and $E_i(\infty)$,  resulting in an estimate of $E_i(\infty)=0.36313(7)$ which closely matches the exact $E_{corr}=0.36320646$ from MOLPRO/MRCC.  Fits for fewer data points are denoted $E_i(\infty)$ in the legend, and included data points up to the value of $N_{ex}$ plotted, for  varying $p, q,$ and $E_i(\infty)$, as well as varying $p$ and $E_i(\infty)$ and fixing $q=-1$, giving $E_i(\infty)=0.36317(4)$.
The shoulder for this system is at 2.3(3)$\times10^5$ excips. }
\label{fig-Ne-V6Z-x3}
\end{figure}

In previous studies of the initiator approximation for FCIQMC, the initiator error has been seen to decay as total particle number increases, but there has been little known about the form of this decay.
In \cref{sec:AppExtrap}, we show that the initiator error is expected to fall off as the inverse of the number of particles in iFCIQMC, and we shall use a similar form to extrapolate results for iCCMC, writing the energy as a function of number of excips, $N_{ex}$:
$$
E_i(N_{ex})=pN_{ex}^q+E_i(\infty),$$
where we will allow $p, q,$ and $E_i(\infty)$ to be variable parameters.
To test the form, we have performed CCSDT calculations on neon in a cc-pV6Z basis set, where the shoulder lies at 2.3(3)$\times10^5$ excips out of a total space of 1.64$\times10^7$ excitors, and we have been able to perform the exact CCSDT calculation with MOLPRO\cite{MOLPRO,MOLPRO-WIREs} and MRCC\cite{MRCC}.
For this system, iCCMC calculations were performed for increasing total excip populations, and projected energy estimators and errors determined using an automated blocking script\footnote{This makes use of the PELE suite (see \protect\url{https://github.com/pele-python/pele}), and we will describe details of this elsewhere}.  As noted in \cref{sec:AppShift}, biases are introduced in calculations in which the shift varies greatly, and we have kept only calculations in which the standard deviation of the shift is lower than $0.05E_h$.
For each population, projected energy estimators for that and lower populations were used in a non-linear least squares fit performed with gnuplot\cite{gnuplot}.  In Figure \ref{fig-Ne-V6Z-x3}, we show fits with $p$ and $q$ allowed to vary, as well as fixing $q=-1$.  The latter leads to considerably more stable fitting, with data from $N_{ex}<25000$ providing estimates of $E_i(\infty)$ matching $E_{corr}$  within error bars (albeit large ones), and we have used it as the basis of our later extrapolations.
The least squares fits produce error estimates on the values of the fitted variables and we have used these as is as error estimates for the extrapolated values, but caution that they are likely to be under-estimates of the errors.

\section{The Uniform Electron Gas}
\label{sec:results}
Whilst the neon atom provides a convenient single-reference system to test the scaling of shoulders with basis, there should be little surprise that the CCSDTQ correlation energy is within a milliHartree of the CCSDT even at the cc-pV6Z basis.
In this section, we turn attention to the three dimensional uniform electron gas whose Wigner--Seitz radius, $r_s$ provides a convenient tunable parameter to different regimes of correlation, and which has been little studied with coupled cluster theory.
Without extensive parallelism (which is to be the subject of a future paper), we have kept the system sizes small, concentrating on the closed-shell 14-electron gas in plane wave bases.  We have used a spherical energy cutoff and denote the number of plane waves $M$.
FCIQMC, CCSD and CCSD(T) results are available in such systems,\cite{ShepherdAlavi_12JCP,ShepherdScuseria_14PRL} along with a well-tested methodology to extrapolate to the complete basis set limit\cite{Shepherd_12PRB035111}.

\begin{figure}
\includegraphics[scale=0.5]{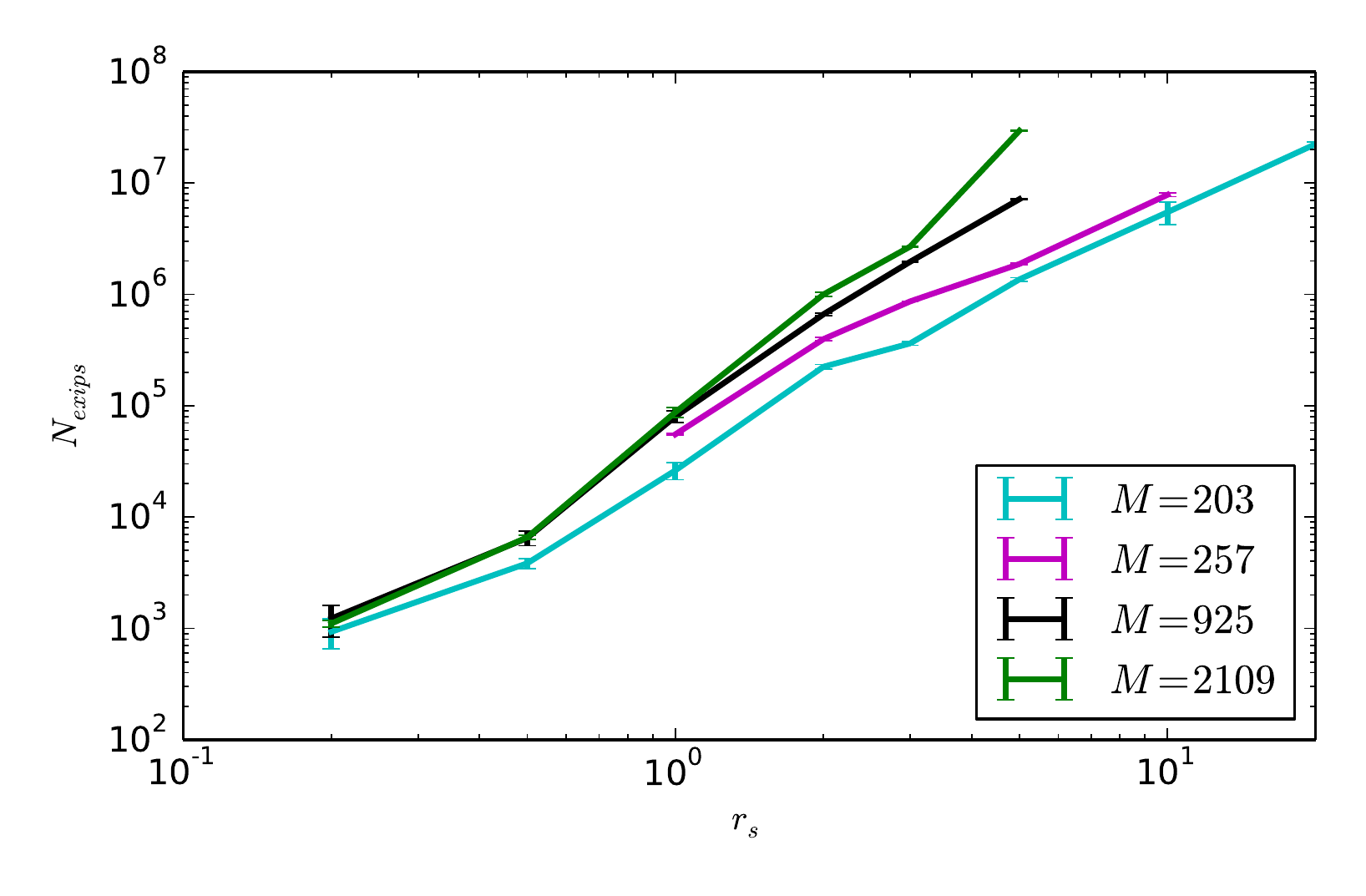}
\caption{The shoulder height for CCSDT on the 3D UEG for systems with $M$ plane waves.}
\label{fig-UEG-plateaux}
\end{figure}

We begin by studying the CCSDT shoulder heights for the UEG as a function of $r_s$.  Figure \ref{fig-UEG-plateaux} shows that these rise significantly with $r_s$, of the order of $r_s^3$, as well as increasing with the plane wave cutoff energy.

\begin{figure}
\includegraphics[scale=0.51]{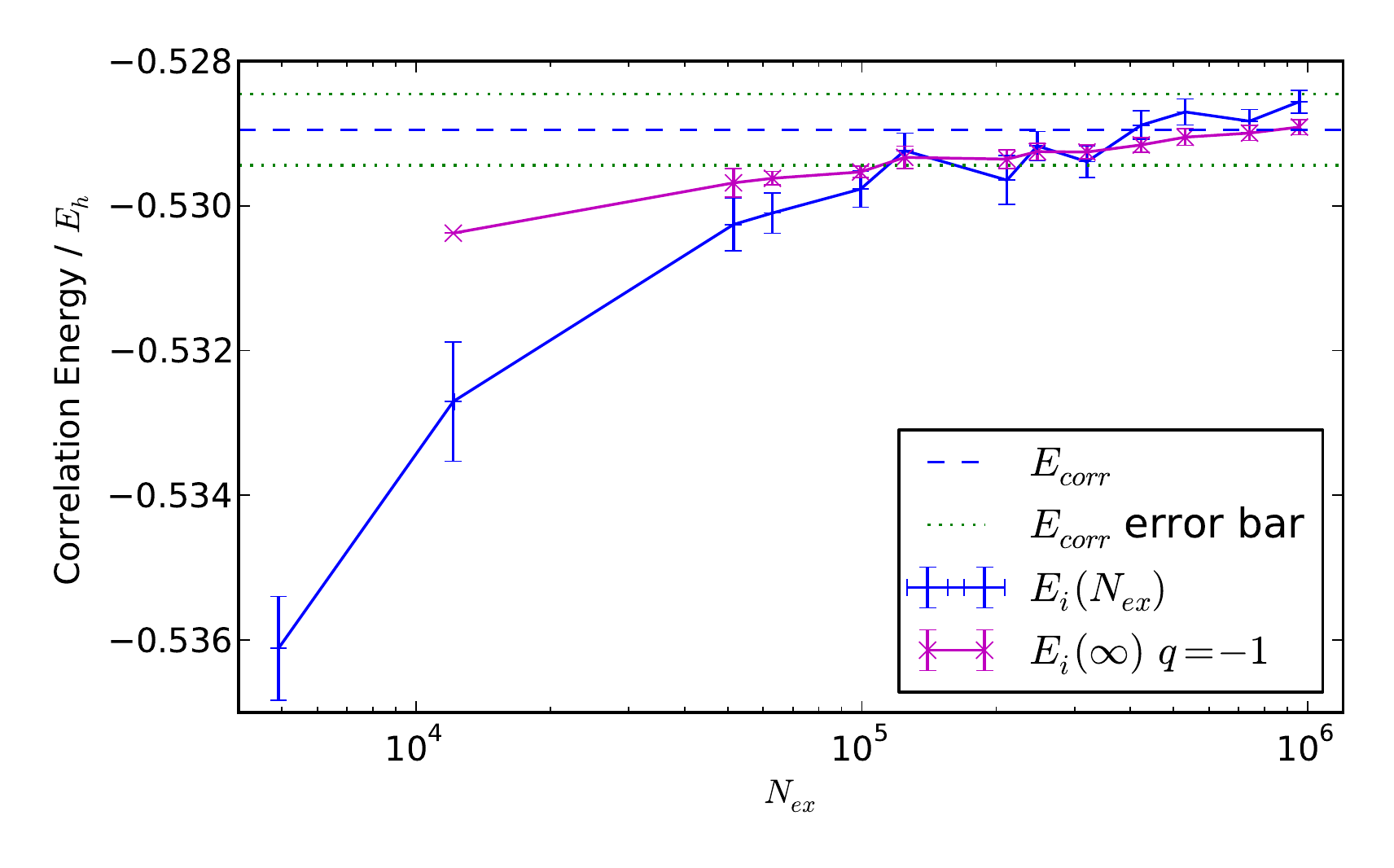}
\caption{CCSDT initiator extrapolation for the 14-electron 3D uniform electron gas with $M=925$ and $r_s=1$.}
\label{fig-UEG-x3-2}
\end{figure}

\begin{figure}
\includegraphics[scale=0.51]{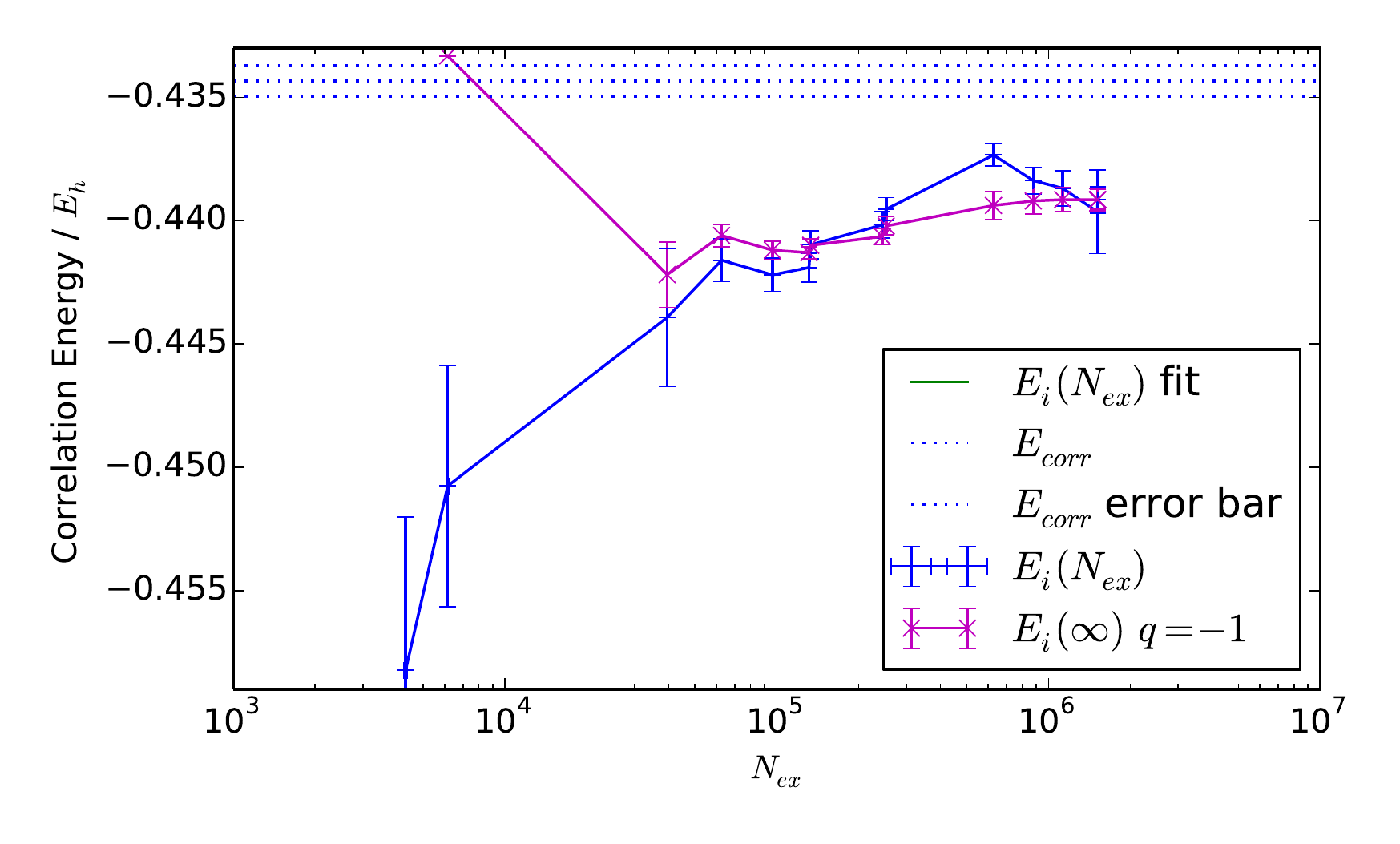}
\caption{CCSDT initiator extrapolation for the 14-electron 3D uniform electron gas with $M=925$ and $r_s=2$. Though the raw initiator energies do not appear to be converging, the extrapolation does appear to converge to a lower energy than the true correlation energy, $E_{corr}$, which was calculated with non-initiator CCSDTMC.}
\label{fig-UEG-x3}
\end{figure}

For values of $r_s\le2$ initiator CCSD and CCSDT calculations were performed followed by linear fits against $1/M$ for $M\ge179$, and the $M=\infty$ limit extracted.
Results are in Table \ref{tab-14eUEG}.
We have also investigated the convergence of the initiator method for these systems, as shown in Figures \ref{fig-UEG-x3-2} and \ref{fig-UEG-x3}.  Whilst the convergence of the extrapolation for $r_s=1$ in Figure \ref{fig-UEG-x3-2} appears to well-match the exact result, we find the same methodology predicts a result many standard deviations away from the true correlation energy for $r_s=2$ (Figure \ref{fig-UEG-x3}).
This non-convergent behaviour of the initiator approximation may be owing to changes in the nature of the correlation with increasing $r_s$, as the UEG is known to become more multireference with increasing $r_s$.  We therefore would caution against the use of initiator extrapolation in such cases where there is not a clear sign of convergence.

\begin{table}
\begin{tabular}{c|ccc}
$r_s$&$E\sr{CCSD}$&$E\sr{CCSDT}$&$E\sr{FCIQMC}$\\
\hline
0.1 & -0.6639(1) & -0.6659(2) &             \\
0.5 & -0.5897(1) & -0.5965(2) & -0.5969(3)  \\
1.0 & -0.5155(3)$^\dagger$ & -0.5317(3) & -0.5325(4)  \\
2.0 & -0.4094(1) & -0.4354(4)$^\ddagger$ & -0.4447(4)  \\
\end{tabular}
\caption{Correlation energies for the 14-electron Uniform Electron Gas. Extrapolated initiator stochastic CC results were obtained by basis-set extrapolation from stochastic coupled cluster calculations. FCIQMC results are from Ref \onlinecite{ShepherdAlavi_12JCP}. $^\dagger$ For $r_s=1$, Shepherd \textit{et al.} have found $E\sr{CCSD}=-0.5152(5)$ (Fig 7 of Ref. \onlinecite{Shepherd_12PRB035111}). $^\ddagger$ Owing to non-converging initiator energies, non-initiator stochastic CC energies were used in the basis set extrapolation.   }
\label{tab-14eUEG}
\end{table}

\section{Conclusion}
\label{sec:conclusion}
We have investigated the stability of Stochastic Coupled Cluster calculations and suggest that the particle ratio (between the total excip population and the population on the reference) is a convenient measure of the progress of a Stochastic Coupled Cluster calculation.
The particle ratio exhibits a maximum as a calculation progresses through the growth phase, and this maximum signals the critical number of particles required for a stable calculation, and indicates a `shoulder height' equivalent to the plateau height in FCIQMC.
As a fraction of the truncated Hilbert space of a coupled cluster calculation, this effective plateau height remains a similar order of magnitude (1--10\%) as basis set is increased, and decreases as the truncation level increases, so allowing an estimate of the upper bound storage required of a CCMC calculation.

We have investigated two possible implementations of the initiator approximation in CCMC, and of the most stable, we conclude that the initiator approximation performs a similar function to that in FCIQMC, and stabilizes calculations with low numbers of excips at the expense of introducing a systematic initiator error.
We have demonstrated that this error, which in all cases we have investigated causes a systematic lowering of the energy, decays as $N\sr{excip}^{-1}$ and this can be used as a basis for extrapolation of the true correlation energy.
Unlike FCIQMC, we have found that the initiator error does not decay away quickly in comparison to the critical stable population indicated by `shoulder plots'.
In both the 3D Uniform Electron Gas and Ne atoms in large bases, we show that the particle population require to reduce the initiator error sufficiently is of the order of the shoulder height.
Therefore the reductions in particle population, which in FCIQMC can amount to many orders of magnitude, do not appear to be present in Stochastic Coupled Cluster Theory.

However, the benefits of the initiator approximation are still manifold. With careful extrapolation, it is possible to perform calculations which are not possible without use of the initiator approximation.
Additionally, the initiator approximation dramatically reduces the variance of the projected energy within a calculation, and thus by its use the required calculation times are greatly reduced.
There is reason to be optimistic as there is much which can be tuned in the Stochastic Coupled Cluster algorithm.  In particular, the role of the population of particles on the reference, which acts as a normalization, is critical in the stability of calculations, and in the projected energy.
With further algorithmic modifications and optimization of parameters, the convergence of the initiator energy with particle number will likely become faster allowing yet larger calculations to be performed.

Finally we demonstrate that this method is capable of calculating coupled cluster energies of the Uniform Electron Gas up to $r_s=2$.  Studies beyond this will require significantly more computational resources, but are still possible.
Of particular note is the vital contribution of triple excitations even at these modest values of $r_s$, and we hope to make a more detailed study of this in the future.  

\begin{acknowledgments}
We are grateful to Miss Ruth Franklin and Mr William Vigor for discussions on the algorithms. AJWT thanks the Royal Society for a University Research Fellowship and Imperial College London for a Junior Research Fellowship, where this work was started.    JSS acknowledges the research environment provided by the Thomas Young Centre under Grant No.~TYC-101. Calculations were performed on the Imperial College High Performance Computing facilities\cite{ICHPC} and the University of Cambridge High Performance Computing facilities using the NECI\cite{BoothAlavi_14MP} and the HANDE\cite{HANDE,JORS15} codes.  Molecular Orbital integrals were generated with a modified version of Q-Chem\cite{QChem}, and for Ne cc-pV6Z, MOLPRO\cite{MOLPRO}.  Raw and analysed data is available in Ref.~\onlinecite{data_dump}.
We use an automated iterative blocking algorithm\cite{FlyvbjergPetersen_89JCP,Wolff2004143,PhysRevE.83.066706,pyblock} to accurately estimate the stochastic error in all CCMC calculations presented in this paper.
 Figures were plotted using matplotlib\cite{Hunter:2007}.
\end{acknowledgments}

\appendix
\section{Initiator Extrapolation}
\label[secinapp]{sec:AppExtrap}
We derive a form for the convergence of the initiator approximation with psip population in Full Configuration Interaction Quantum Monte Carlo.
We expect that, while CCMC behaviour will be more complicated, it should follow the same general form.

Let us denote the true ground state wavefunction $\psi_0$ with energy $E_0$ and impose intermediate normalization on it such that $\brket{\D{0}}{\psi_0}=1$.
Consider the instantaneous wavefunction at a given timestep, $\psi(\tau)$.
We can expect it to consist of a some amount of the ground state, and a difference term, $\psi(\tau)=C(\tau)\psi_0+\Delta(\tau),$ where we can specify that $\brket{\D{0}}{\Delta(\tau)}=0.$
In a converged simulation $C$ is the population of the reference determinant, and is proportional to the total number of particles.
The projected energy is given by
\begin{equation}
E(\tau)=\frac{\braket{\D{0}}{\hat{H}}{\psi(\tau)}}{\brket{\D{0}}{\psi(\tau)}}=\frac{\braket{\D{0}}{\hat{H}}{\psi(\tau)}}{C(\tau)}.
\end{equation}

In an unbiased FCIQMC calculation $\bket{\Delta(\tau)}_\tau=0$ (where $\bket{}_\tau$ indicates the long time average over $\tau$).
In moving to the initiator modification, we introduce a systematic bias in the dynamics, resulting in a $\bket{\Delta(\tau)}_\tau$ which no longer vanishes.
For a given value of the initiator threshold and value of $C$, there will be a maximum component of $\Delta(\tau)$, i.e. some $\delta$, such that for all determinants $\bf i$, $\brket{\D{i}}{\Delta(\tau)}<\delta$.
Let us assume that any dependence $\delta$ has on $C$ is at most of order $C^0$ (i.e. constant), as a higher power would lead to the initiator approximation not converging with increasing psip population.

Explicitly evaluating the projected energy,
\begin{eqnarray}
E(\tau)&=&\frac{\braket{\D{0}}{\hat{H}}{C(\tau)\psi_0+\Delta(\tau)}}{C(\tau)}\\
&=&C(\tau)\frac{\braket{\D{0}}{\hat{H}}{\psi_0}}{C(\tau)}+\frac{\sum_{\bf i\ne\bf 0}\braket{\D{0}}{\hat{H}}{\D{i}}\brket{\D{i}}{\Delta(\tau)}}{C(\tau)}\\
&\le&E_0+\frac{\sum_{\bf i}H_{0\bf i}\delta}{C(\tau)}.
\end{eqnarray}
Denoting the maximal value of $H_{0\bf i}$ as $\eta$, if there are $N\sr{doub}$ double excitations of the reference, we find
\begin{eqnarray}
|E(\tau)-E_0|&\le&\frac{N\sr{doub}\eta\delta}{C(\tau)}.
\end{eqnarray}
Since $N\sr{doub}$ is constant, and $C(\tau)$ scales linearly with $N\sr{psip}$, we may attempt to fit to $(N\sr{psip})^{-1}$ as an upper bound.

\section{Population Control Effects}
\label[secinapp]{sec:AppShift}

\Cref{sec:AppExtrap} contains the assumption that the initiator approximation is the cause of all systematic errors in the simulation.  This is not strictly true; the effect of population control is well understood to introduce a small bias which decays with increasing population\cite{UmrigarRunge_93JCP,VigorThom_15JCP}.  We briefly investigate here the effect of population control on the convergence of the initiator error.

\begin{figure}
    \includegraphics[scale=0.51]{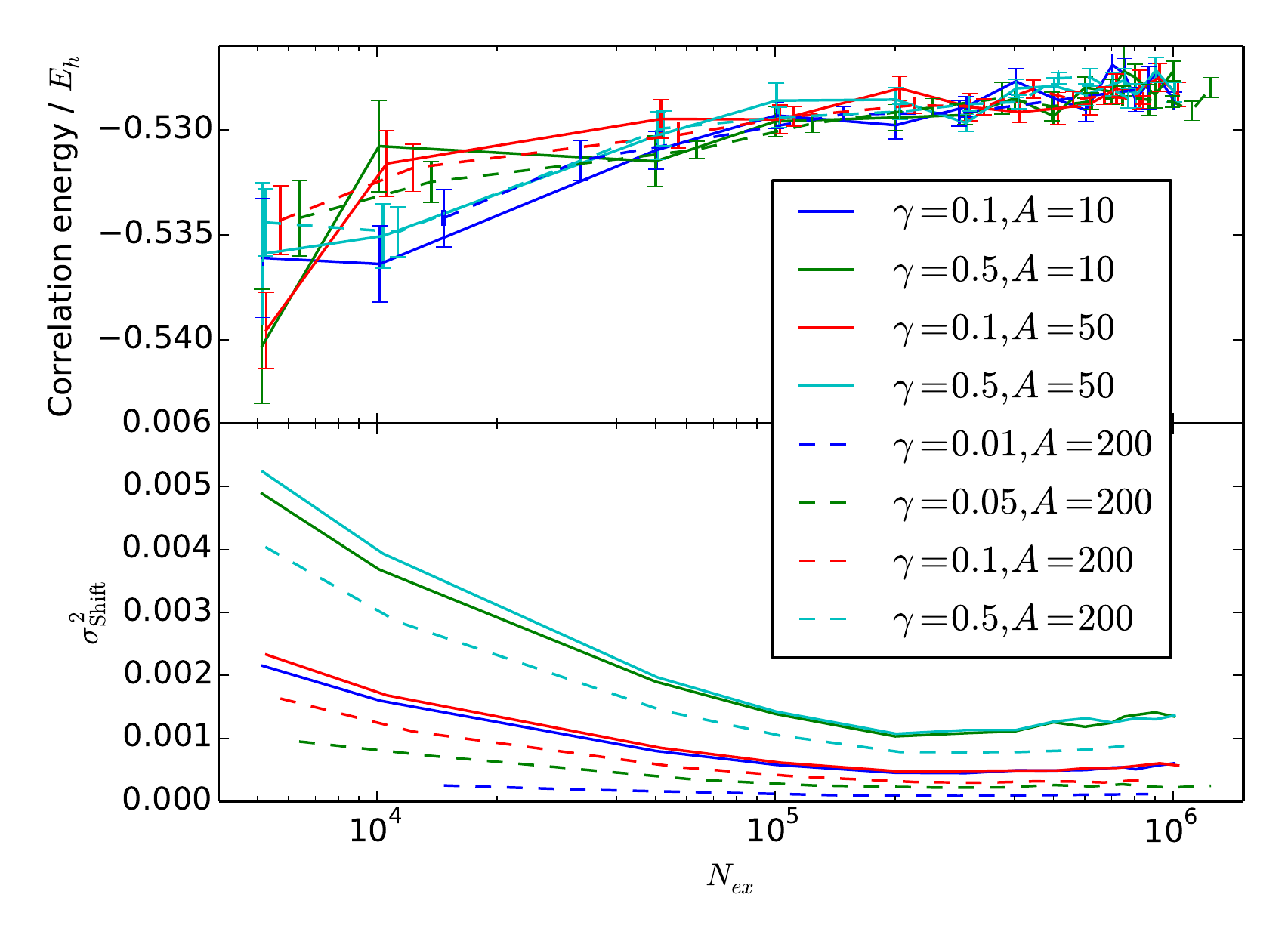}
    \caption{iCCMC convergence studies of CCSDT for the 14-electron 3D uniform electron gas at $r_s=1$ using $M=925$ plane waves for different population control parameters (top) and the corresponding variance in the shift observed during each calculation (bottom).  $\delta\tau=10^{-4}$ in all calculations.}
    \label{fig-UEG-x3-shift}
\end{figure}

\begin{figure}
    \includegraphics[scale=0.51]{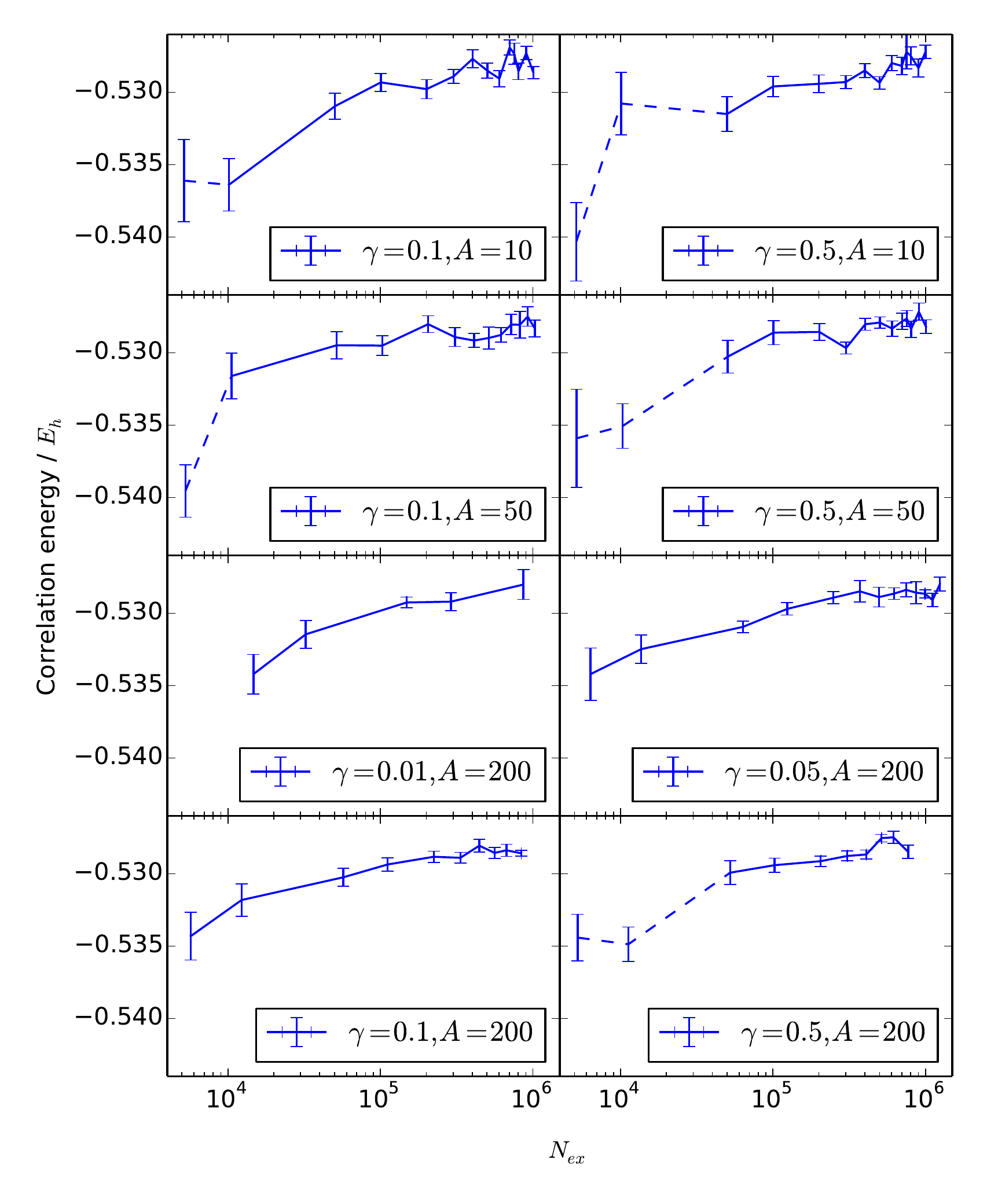}
    \caption{iCCMC convergence studies of CCSDT for the 14-electron 3D uniform electron gas at $r_s=1$ using $M=925$ plane waves for different population control parameters.  $\delta\tau=10^{-4}$ in all calculations.  Dashed lines represent calculations for which the shift variance is greater than 0.002. This is a more detailed view of the top panel in \cref{fig-UEG-x3-shift}.}
    \label{fig-UEG-x3-shift-split}
\end{figure}

The shift is updated every $A$ timesteps during the population control phase according to\cite{BoothAlavi_09JCP,UmrigarRunge_93JCP}
\begin{equation}
    S(\tau) = S(\tau-A\delta\tau) - \frac{\gamma}{A\delta\tau} \ln\left( \frac{N_{ex}(\tau)}{N_{ex}(\tau-A\delta\tau)} \right),
\end{equation}
where $\gamma$ is a (typically small) damping parameter.  The variance in the shift (which is independent of the length of a simulation) is a measure of the population fluctuations caused by requiring the population to be approximately constant.  \cref{fig-UEG-x3-shift,fig-UEG-x3-shift-split} show the effect of population control on the 14-electron UEG at $r_s=1$.  All simulations converge to within stochastic error at sufficiently large populations (\cref{fig-UEG-x3-shift}); however extrapolation also depends upon data from simulations using smaller populations of excips.  Whilst this is far from a comprehensive study, we observe a relatively smooth convergence in all cases once $\sigma^2_{\mathrm{Shift}} < 0.002$ (\cref{fig-UEG-x3-shift-split}).  Therefore extrapolation requires careful choices of these parameters.

\bibliography{iCCMC}

\end{document}